\documentclass[english,aps,preprint]{revtex4}
\usepackage[]{fontenc}
\usepackage[latin9]{inputenc}
\usepackage{float}
\usepackage{amstext}
\usepackage{graphicx}
\usepackage{esint}

\makeatletter

\newcommand{\lyxdot}{.}

\@ifundefined{textcolor}{}
{%
 \definecolor{BLACK}{gray}{0}
 \definecolor{WHITE}{gray}{1}
 \definecolor{RED}{rgb}{1,0,0}
 \definecolor{GREEN}{rgb}{0,1,0}
 \definecolor{BLUE}{rgb}{0,0,1}
 \definecolor{CYAN}{cmyk}{1,0,0,0}
 \definecolor{MAGENTA}{cmyk}{0,1,0,0}
 \definecolor{YELLOW}{cmyk}{0,0,1,0}
 }

\@ifundefined{showcaptionsetup}{}{%
 \PassOptionsToPackage{caption=false}{subfig}}
\usepackage{subfig}
\makeatother

\usepackage{babel}

\begin{document}

\title{Black hole phase transitions in Ho\v{r}ava-Lifshitz gravity}

\author{Qiao-Jun Cao}

\email{caoqiaojun@gmail.com}

\author{Yi-Xin Chen}

\email{yxchen@zimp.zju.edu.cn}

\author{Kai-Nan Shao}

\email{shaokn@gmail.com}

\affiliation{Zhejiang Institute of Modern Physics, Zhejiang University, Hangzhou,
310027, China}
\begin{abstract}
We study black hole phase transitions in (deformed) Ho\v{r}ava-Lifshitz
(H-L) gravity, including the charged/uncharged topological black holes
and KS black hole. Stability analysis and state space geometry are
both used. We find interesting phase structures in these black holes,
some of the properties are never observed in Einstein gravity. Particularly,
the stability properties of black holes in H-L gravity with small
radius change dramatically, which can be considered as a leak of information
about the small scale behavior of spacetime. A new black hole local
phase transition in H-L gravity which cannot be revealed by thermodynamical
metrics has been found. There is an infinite discontinuity at the
specific heat curve for charged black hole in H-L gravity with hyperbolic
event horizon. However, this discontinuity does not have a corresponding
curvature singularity of thermodynamical metrics. Our results may
provide new insights towards a better understanding of the H-L gravity,
as well as black hole thermodynamics.

\end{abstract}

\pacs{04.70.Dy, 04.50.-h, 04.90.+e}

\maketitle

\section{introduction\label{sec:introduction}}

Due to the celebrated work of Bekenstein\cite{Bekenstein1973} and
Hawking\cite{Hawking1975} on black hole physics, it is widely accepted
that black holes are thermodynamic systems. The four laws of black
hole thermodynamics\cite{bardeen1973four} have been put on solid
fundamentals. It is also believed that some kinds of unknown microscopic
degrees of freedom are coming into play since every black hole has
a temperature proportional to the surface gravity on the black hole
event horizon and an entropy proportional to the area of event horizon.

Phase transition is an important phenomenon in ordinary thermodynamics.
It is therefore natural to ask whether there are some phase transitions
in black hole thermodynamics. The pioneering work of Hawking and Page
proved that there is a \emph{global} phase transition between thermal
AdS state and AdS black hole in four dimensions as the temperature
changes\cite{hawking1983thermodynamics}. It means that thermal radiation
in AdS space becomes unstable at a certain temperature, and eventually
collapses to form a black hole. However, physical difference in the
microscopic structure in different phases is still unclear. Witten
reconsidered Hawking-Page phase transition in the spirit of AdS/CFT
correspondence, and interpreted it as a transition between low-temperature
confining and  high temperature deconfining phase in the dual field
theory\cite{witten1998anti}. The research of black hole phase transition
has been extended and indicate that there may exist different phase
transitions in various circumstance\cite{davies1989thermodynamic,chamblin1999charged,chamblin1999holography,caldarelli2000thermodynamics,carter2005thermodynamics,cai2002gauss,cai2007ricci,myung2008phase,dey2007phase,sheykhi2009thermodynamical,fernando2006thermodynamics,myung2008thermodynamics,anninos2009thermodynamics,dehghani2009thermodynamic}.
The signal of a phase transition is typically characterized by the
sign change of a capacity, such as the specific heat, charge capacitance
or moment of inertia, by going through either zero or infinity. Davies
argued that the points that the specific heat with fixed charge and/or
angular momentum passing from negative to positive values through
an infinite discontinuity are commonly associated with phase transitions\cite{davies1977thermodynamic,davies1978thermodynamics}.
These divergences in the phase space are denoted as Davies curves
now.  We should also note that this type of thermodynamical phase
transition is based on the local stability analyses and is generally
different from the global Hawking-Page phase transition. In this paper,
we shall keep an eye to distinguish the local and global stabilities
of black hole thermodynamic ensembles.

The study of phase transition has been extended in various directions,
e.g., to black holes with non-spherical event horizon topologies\cite{birmingham1999topological,koutsoumbas2008phase},
to black holes including more conserved quantities such as charges
and angular momenta\cite{davies1989thermodynamic,chamblin1999charged,chamblin1999holography,caldarelli2000thermodynamics,carter2005thermodynamics}
or to higher dimensional black holes\cite{carter2005thermodynamics}.
The research can also be generalized to gravity theories with high
derivatives of curvature terms, such as Gauss-Bonnet gravity\cite{cai2002gauss,cai2007ricci,anninos2009thermodynamics},
Born-Infeld gravity\cite{myung2008thermodynamics,fernando2006thermodynamics,sheykhi2009thermodynamical}
and Lovelock gravity etc\cite{myung2008phase,dey2007phase,dehghani2009thermodynamic}.
In these cases, phase transitions and critical phenomena are more
complicated.

Weinhold\cite{weinhold1975} and Ruppeiner\cite{ruppeiner1979thermodynamics,ruppeiner1995riemannian}
introduced thermodynamical metrics to study thermodynamic systems.
A thermodynamic system is described by some thermodynamical quantities,
and these quantities are all related to each other. Weinhold took
internal energy $M$ as a function of entropy $S$ and other extensive
variables $N^{a}$ which are necessary to describe the thermodynamic
system. Denoting the collection of all the other thermodynamical quantities
except $M$ as $\left\{ S^{i}\right\} \equiv\{S,N^{a}\}$, then $M=f(S^{i})$.
He constructed a metric $ds^{2}(M)=h_{ij}dS^{i}dS^{j}$, where $h_{ij}=\partial^{2}f(S^{i})/(\partial S^{i}\partial S^{j})$
is the Hessian matrix of $f(S^{i})$. In the similar way, Ruppeiner
later introduced a new metric $ds^{2}(S)$ by using the entropy as
a generating function depending on other extensive variables\cite{ruppeiner1979thermodynamics}.
Remarkably, the curvature of Ruppeiner metric could be related to
interactions in the underlying statistical system and scaled as the
correlation volume. It is divergent at critical points of second order
phase transitions. It has been proved that Weinhold and Ruppeiner
metrics are conformally equivalent to the inverse of the temperature
as the conformal factor. There are amounts of works in discussing
the relationship of black hole phase transitions and Ruppeiner/Weinhold
metrics\cite{aman2003geometry,shen2007thermodynamic,ruppeiner2008thermodynamic,sahay2010thermodynamic}.
The Ruppeiner and Weinhold metrics indeed reveal some signals of black
hole phase transitions, but sometimes they fail to work to the Davies
curve\cite{ruppeiner2008thermodynamic}.

Inspired by Weinhold and Ruppeiner, a framework was proposed to identify
all of the phase transition points in black hole thermodynamics\cite{liu2010thermodynamical}.
The authors of \cite{liu2010thermodynamical} introduced metrics based
on all thermodynamical potentials generated by Legendre transformations
of the black hole energy or entropy, and found that the collection
of all phase transition points are in one-to-one correspondence to
the collection of curvature singularities of those thermodynamical
metrics. This work clarified that the divergences in the associated
set of thermodynamical geometries can reveal the threshold points
of all capacities completely. A natural question is {}``what is range
of application of the correspondence between curvature singularities
of the generalized thermodynamical metrics and phase transition points
in black hole thermodynamics?'' or {}``Is the correspondence valid
for all thermodynamical systems? e.g. black hole solutions among different
gravity theories such as Born-Infeld gravity and Lovelock gravity
etc.'' Though various black holes with spherical topology of horizon
and black ring in five dimensions with single angular momentum were
investigated in \cite{liu2010thermodynamical}, it is far from a complete
proof of the correspondence between curvature singularities of thermodynamical
metrics and phase transition signals. It will be very interesting
to investigate some black holes in other gravity theories or with
non-spherical topology of horizon that may provide counter-examples
to this correspondence, which may help to identify the correspondence's
range of application.

Recently, a new theory of gravity at a Lifshitz point was proposed
by Ho\v{r}ava, which may be regarded as a UV complete candidate for
general relativity\cite{Horava2009,Horava2009a,Horava2009b}. The
Ho\v{r}ava-Lifshitz (H-L) theory has been intensively investigated.
Some kind of black hole solutions for H-L gravity have been found\cite{Lu2009a,cai2009topological,kehagias2009black,Ghodsi:2009zi}.
The thermodynamic analysis of H-L gravity black holes is a nontrivial
task and has been done in \cite{cai2009topological,cai2009thermodynamics}.
Nevertheless, the charged black hole in H-L gravity has not been studied
from the thermodynamic perspective. Moreover, in \cite{biswas2010geometry}
and \cite{wei2010thermodynamic}, the authors examined the thermodynamical
geometry of the topological black hole in H-L theory and KS black
hole, respectively. Their papers only constructed the Ruppeiner and
Weinhold metrics. However, the new metric based on free energy which
was introduced in \cite{liu2010thermodynamical}should be included
for a full analysis. These prompt us to study the  stability of black
holes and their phase transitions in H-L gravity. It is also necessary
to construct their whole set of thermodynamical metrics and calculate
the associated Ricci scalars, and to examine the correspondence mentioned
in the previous paragraph.

In this paper, we investigate possible black hole phase transitions
in H-L gravity carefully, including the charged/uncharged topological
black hole in H-L gravity and KS black hole. We discuss stability
of these black holes as well as the phase transition signals using
both analytical and graphical techniques, especially in the ensemble
with fixed charge parameters. The phase structures are vivid in these
kinds of black holes in H-L gravity, which are very different from
the black holes in Einstein gravity. Thermodynamic properties of the
topological black holes in H-L gravity are reexamined, and new observations
have been made. We analyze the specific case with dynamical coupling
constant $\lambda=1$ carefully, for black holes with hyperboloid
topology. It turns out that there is a phase transition going from
small to large black holes at a critical temperature, of first order.
For charged black hole in H-L gravity, in the ensemble with fixed
charge parameters, we found that the black holes with sphere or flat
topology have no phase transition in the canonical ensemble with fixed
charge, since the specific heat with fixed charge $C_{Q}$ is positive
definite. However, the thermodynamic behavor of RN AdS black holes
is different, there can be three branches of RN AdS black holes with
a small fixed charge, and the middle radius branch is unstable, with
the temperature increases, small black holes can phase transition
to large black holes\cite{chamblin1999charged,chamblin1999holography}.
The charged black hole in H-L gravity with hyperboloid topology is
extremely interesting. There are three branches of black holes. The
branch with middle radius is local unstable while the small and large
branches are local stable. This signifies a local phase transition
at the point with divergent specific heat, but there is no global
phase transition here since the branch with largest radius always
have minimum free energy at all temperatures. For KS black hole, it
is obvious to see that the high derivative terms in the H-L gravity
action play an important role. Their influences on the KS black hole
thermodynamic properties reduced to a charge like parameter, and make
it thermodynamically behaving like a RN black hole. Compared to the
Schwarzschild black hole it has a new stable phase with a small radii,
and phase transition comes up.

The thermodynamical metrics of black holes in H-L gravity are also
constructed, and the corresponding Ricci scalars are calculated for
the purpose of investigating their curvature singularities. We found
that all the cases we discussed in this paper are almost consistent
with the framework proposed in \cite{liu2010thermodynamical}, except
that there is a infinite discontinuity at the specific heat curve
for charged black hole with hyperbolic event horizon in H-L gravity,
while this discontinuity does not has a corresponding curvature singularities
of thermodynamical metrics. This is a probable counter-example to
the correspondence between local phase transition points and curvature
divergences of thermodynamical geometries. This also indicate that
the thermodynamical metrics can not reveal all of the local phase
transition signals in H-L gravity. The violation of this correspondence
in charged black holes with hyperbolic event horizon in H-L gravity
may be related to the non-spherical topology of it's event horizon.
The ultraviolet behavior of spacetime in H-L gravity deserves as the
reason for all of those strange properties in black hole thermodynamics.
Our results may provide new insights towards a better understanding
of the Ho\v{r}ava-Lifshitz gravity, as well as black hole thermodynamics.

This paper is organized as follows. In Section \ref{sec:Topological-black-hole},
we focus on the thermodynamic properties of topological black holes
in H-L gravity, and recall the stability analyses by using their heat
capacity and free energy. In Section \ref{sec:Charged-black-hole},
we consider the charged topological black holes in H-L gravity. Their
 stability and the phase transition signals are examined carefully.
Both the specific heats $C_{Q}$ and $C_{\Phi}$, and charge capacitances
$\tilde{C_{T}}$ and $\tilde{C}_{S}$ are calculated out. The threshold
points where these quantities change sign through zero or infinite
discontinuity are compared with the divergent points of the Ricci
scalar curvatures for Weinhold metric $ds^{2}(M)$, Ruppeiner metric
$ds^{2}(S)$ and the free-energy metric $ds^{2}(F)$. We found that
\emph{not} all threshold points in the four $C$'s match precisely
to those singularities in the three metrics. In Section \ref{sec:KS-black-hole},
the thermodynamical properties of KS black holes were investigated.
It thermodynamically behaves like a RN black hole, and phase transition
comes up. We see that all possible phase transitions found from the
four $C$'s correspond to curvature singularities of certain thermodynamical
metrics, which are consistent very well to the framework proposed
by \cite{liu2010thermodynamical}. Section \ref{sec:Conclusion} is
for summary and discussions. In the appendix, we examine the thermodynamical
properties of KS black holes by defining the mass and the entropy
in an alternative way.

\section{Topological black hole in Ho\v{r}ava-Lifshitz gravity\label{sec:Topological-black-hole}}

Thermodynamic properties of the topological black holes in H-L gravity
have been examined in \cite{cai2009thermodynamics,biswas2010geometry}.
We review it briefly in this section and make some new observations.
For general dynamical coupling constant $\lambda$ in H-L gravity
action, the solution of the topological black holes in H-L gravity
was first obtained in \cite{cai2009topological}. The metric can be
written as \begin{equation}
ds^{2}=-\tilde{N}^{2}f(r)\, dt^{2}+\frac{dr^{2}}{f(r)}+r^{2}d\Omega_{k}^{2}\quad,\label{eq:metric of topological bh}\end{equation}
where $d\Omega_{k}^{2}$ denotes the line element for two-dimensional
Einstein space with constant scalar curvature $2k$. We can take (without
loss of generality) $k=0\,,\pm1$ for plane, sphere or hyperboloid
2-space respectively. The function $f(r)$ is given by\begin{equation}
f(r)=k-\Lambda r^{2}-\alpha r^{s}\quad,\qquad\tilde{N}=\gamma r^{1-2s}\quad,\end{equation}
where\begin{equation}
s=\frac{2\lambda\pm\sqrt{2\left(3\lambda-1\right)}}{\lambda-1}\quad,\label{eq:s}\end{equation}
and $\alpha$ and $\gamma$ are both integration constants. The black
hole horizon is at $r=r_{+}$, which is defined by the largest real
positive root of $f(r)=0$. There are two branches of solution according
to the sign in (\ref{eq:s}). It is reasonable to choose the negative
branch since the physical meaning of the positive branch is not very
clear\cite{cai2009thermodynamics}. The range of $s$ is $\left(-1,\,2\right)$
for the negative branch when $\lambda>1/3$.

It should be noticed that the asymptotically behavior of those solutions
are complicated. They are neither asymptotically flat nor asymptotically
AdS, so we have to use the canonical Hamiltonian method to define
their mass. Defining $l^{2}=-\frac{1}{\Lambda}$, the thermodynamical
quantities are collected as\cite{cai2009thermodynamics}\begin{eqnarray}
T & = & \frac{\gamma}{8\pi r_{+}}\left(-\Lambda r_{+}^{2}\left(2-s\right)-s\, k\right)=\frac{\gamma}{4\pi r_{+}^{2s}}\left[\left(\frac{r_{+}}{l}\right)^{2}\left(2-s\right)-k\, s\right]\quad,\label{eq:topoT}\\
S & = & \frac{\pi\kappa^{2}\mu^{2}\Omega_{k}}{\sqrt{2\left(3\lambda-1\right)}}\left[k\ln\left(\sqrt{-\Lambda}r_{+}\right)+\frac{1}{2}\left(\sqrt{-\Lambda}r_{+}\right)^{2}\right]+S_{0}\quad,\\
M & = & \frac{\sqrt{2}\kappa^{2}\mu^{2}\Omega_{k}}{16\sqrt{3\lambda-1}}\gamma\alpha^{2}=\frac{\sqrt{2}\kappa^{2}\mu^{2}\Omega_{k}\gamma}{16\sqrt{3\lambda-1}}\frac{\left(k-\Lambda r_{+}\right)^{2}}{r_{+}^{2s}}\quad.\end{eqnarray}
 Using\begin{equation}
c=\frac{\kappa^{2}\mu}{4}\sqrt{\frac{\Lambda}{1-3\lambda}}=\left(\frac{2-s}{1+s}\right)\left(\frac{\kappa^{2}\mu}{4\sqrt{2}l}\right)\quad,\end{equation}
one may rewrite\begin{equation}
M=\frac{c^{3}}{16\pi G}\left(\frac{1+s}{2-s}\right)\left(\gamma\Omega_{k}l^{2-2s}\right)\left[\frac{k+\left(\frac{r_{+}}{l}\right)^{2}}{\left(\frac{r_{+}}{l}\right)^{s}}\right]^{2}\quad,\label{eq:TopoMass}\end{equation}
\[
S=\frac{c^{3}}{4G}\left(\frac{1+s}{2-s}\right)\left(\Omega_{k}l^{2}\right)\left[k\ln\left(\frac{r_{+}}{l}\right)^{2}+\left(\frac{r_{+}}{l}\right)^{2}\right]+S_{0}\quad.\]
It is easy to confirm that the above thermodynamical quantities satisfy
the first law of thermodynamics $dM=TdS$.

Two points should be noticed here which did not appear in the analyses
of Schwarzschild AdS black holes in \cite{hawking1983thermodynamics},
viz., the expression of mass (\ref{eq:TopoMass}) is not monotonous
increasing with $r_{+}$ when $k=\pm1$, and there is a logarithmic
term in the entropy expression unless $k=0$, one cannot fix the integration
constant $S_{0}$ here.  The sign of the specific heat for the topological
black holes in H-L gravity is different from the sign of local slope
of the $T(r_{+})$ curve. We should be careful about them in the following
discussion.

\begin{figure}[H]
\begin{centering}
\subfloat[$k=1$]{\includegraphics[scale=0.66]{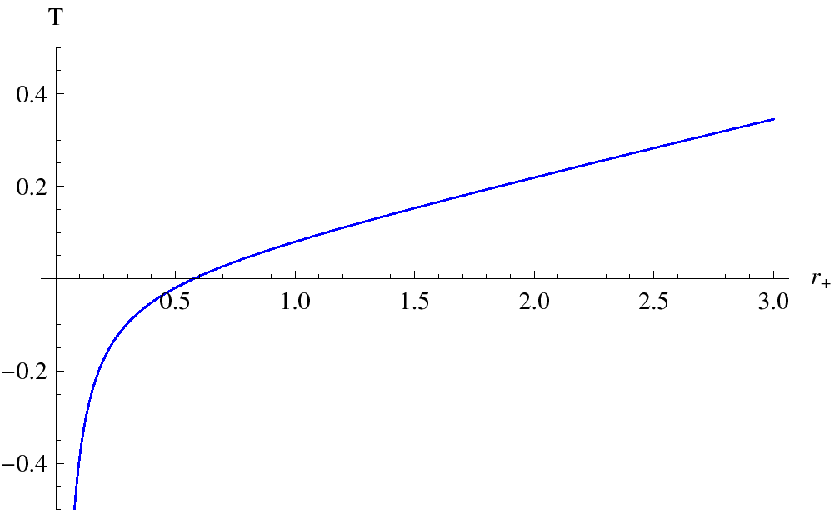}}\subfloat[$k=0$]{\includegraphics[scale=0.66]{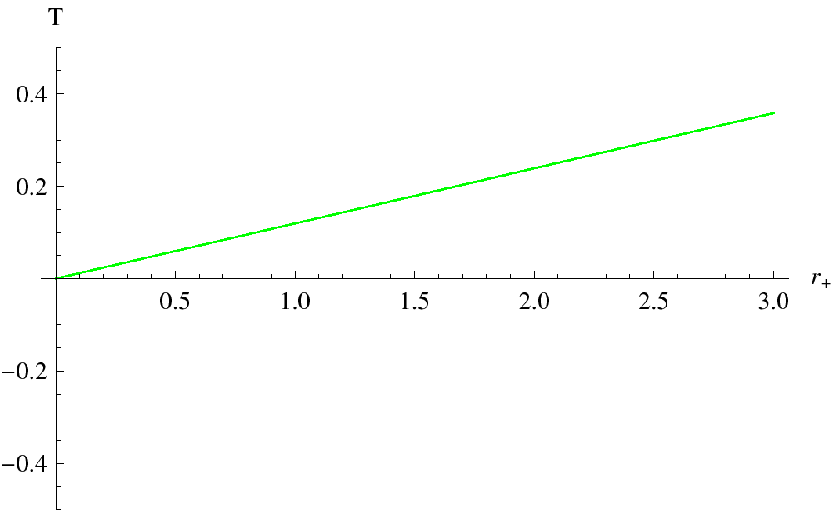}}
\par\end{centering}

\centering{}\subfloat[$k=-1$ \label{fig:topoTm1}]{\includegraphics[scale=0.7]{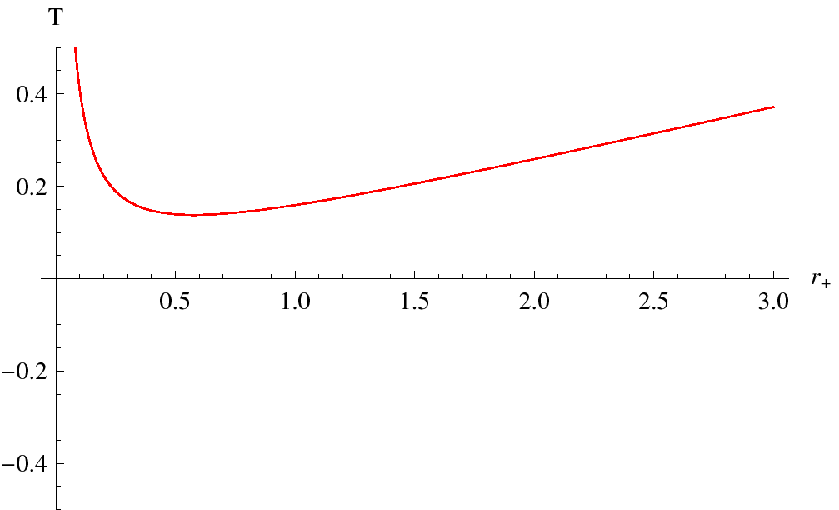}}\caption{Temperature for topological black holes \emph{vs}. horizon radius
$r_{+}$, with $c=l=G=\gamma=\Omega_{k}=1$ and $s=\frac{1}{2}$.
For $k=1$, the negative temperature region $r_{+}<\frac{1}{\sqrt{3}}$
is unphysical, only the positive temperature region have physical
significance, the zero temperature point is where the black hole is
extremal. In the case of $k=-1$, the behavior of the temperature
is similar to Schwarzschild AdS black hole. (see text for discussion
of this case.)\label{fig:Temperature-topo}}

\end{figure}

\begin{figure}[H]
\centering{}\includegraphics[scale=0.8]{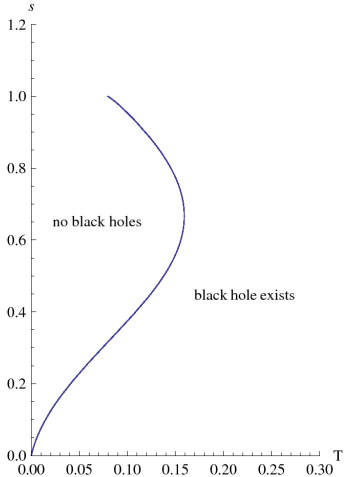}\caption{Phase structure for topological black holes with $k=-1$ and $0<s<1$.
Black holes only exists in the right side of the curve of minimum
temperature, i.e. above a certain temperature.\label{fig:topoPhase}}

\end{figure}

It is instructive to plot the temperature as a function of horizon
radius for the topological black holes in H-L gravity (see Figure
\ref{fig:Temperature-topo}). In these figures, we have taken the
parameters $c=l=G=\gamma=\Omega_{k}=1$ and $s=\frac{1}{2}$ ($\lambda=1$).
As can be seen from Figure \ref{fig:topoTm1}, the temperature for
$k=-1$ has a minimum, which locates at $r_{+}=\frac{s}{\sqrt{2-3s+s^{2}}}$.
Above the minimum temperature, there are two black holes with different
radius associated with the same temperature. Below this temperature,
no black hole exists. Those facts may indicate that there is a phase
transition at this point. Checking the expression of temperature Eq.(\ref{eq:topoT}),
one can see that the minimum temperature only exists for $0<s<1$.
Figure \ref{fig:topoPhase} depicts the phase structure for topological
black holes with $k=-1$. As shown in Figure \ref{fig:topoPhase},
those black holes only exists above a minimum temperature. This is
qualitatively similar to the familiar case of the Schwarzschild AdS
black holes studied in \cite{hawking1983thermodynamics}. It should
be emphasized that this is not all the story of the black holes discussed
here. This is because the expressions of mass and entropy are not
monotonous increase with horizon radius when $k=-1$, and we cannot
conclude that the black hole with smaller radii is unstable while
the larger one is stable. To be more precise, the sign of the specific
heat $C=\frac{\partial M}{\partial T}=T\frac{\partial S}{\partial T}$,
which determines the local stability of a thermodynamic system, cannot
be inferred from the local slope of the $T(r_{+})$ curve for the
topological black holes in H-L gravity. This is very different from
the Hawking-Page transition. The local stability properties of those
black holes should be analyzed by using the specific expression of
specific heat.

It is also worth to mention another interesting issue that the temperature
behaviors of the topological black holes in H-L gravity with $k=1$,
$0$ and $-1$ are very similar to the black holes with $k=-1$, $0$
and $1$ in Einstein's general relativity, respectively, which has
been first emphasized in \cite{cai2009topological}.

\begin{figure}[H]
\begin{centering}
\subfloat[$k=1$]{\centering{}\includegraphics[scale=0.66]{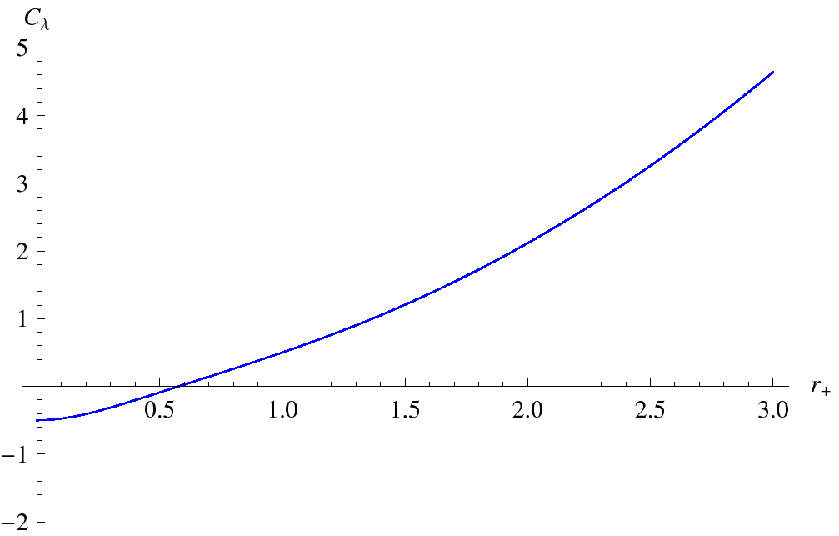}}\subfloat[$k=0$]{\centering{}\includegraphics[scale=0.66]{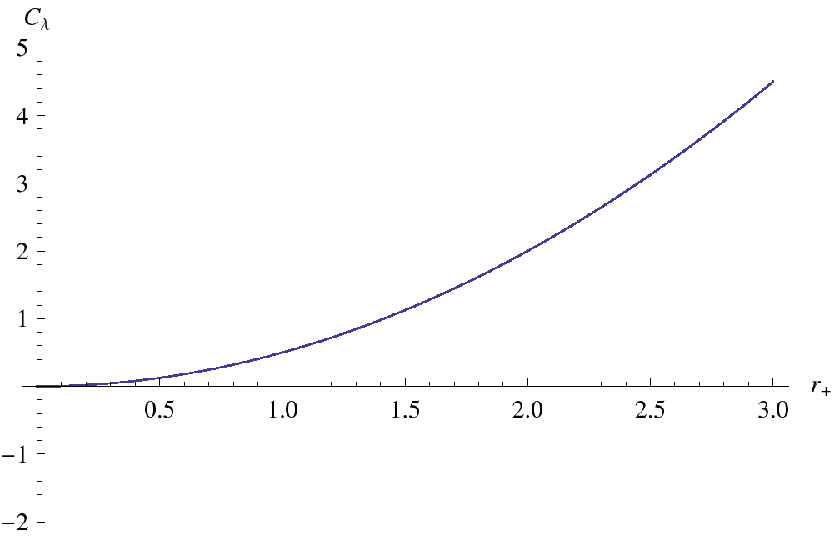}}
\par\end{centering}

\centering{}\subfloat[$k=-1$\label{fig:TopoCkm1}]{\centering{}\includegraphics[scale=0.7]{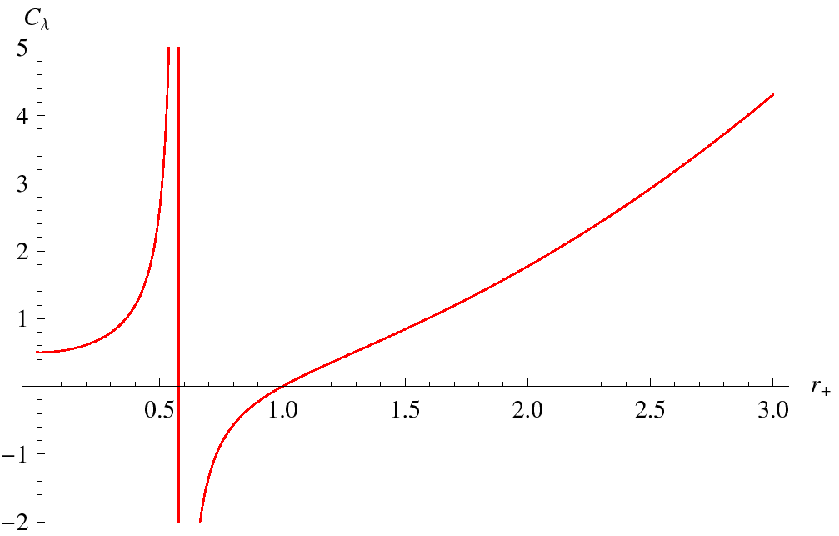}}\caption{Heat capacity for topological black holes \emph{vs}. horizon radii
$r_{+}$, with $c=l=G=\gamma=\Omega_{k}=1$ and $s=\frac{1}{2}$.
For $k=1$, the heat capacity changes from negative values to positive
values through a zero point has nothing to do with phase transition,
the negative region $r_{+}<\frac{1}{\sqrt{3}}$ is unphysical, and
the zero point is associated with a extremal black hole. The divergence
in the last graph is associate with the minimum temperature in Figure
\ref{fig:topoTm1}. There are three branches of black holes, the middle
branch is local unstable with negative values of heat capacity while
the small and large branches are local stable. \label{fig:TopoC}}

\end{figure}

To study the local stability of topological black holes in H-L gravity,
we need to calculate the heat capacity, \begin{equation}
C_{\lambda}=\frac{\partial M}{\partial T}\biggl|_{\lambda}=\frac{c^{3}l^{-2s}r^{2s}\left(\frac{r}{l}\right)^{-2s}\left(kl^{2}+r^{2}\right)(1+s)\left(r^{2}(-2+s)+kl^{2}s\right)\text{\ensuremath{\Omega_{k}}}}{4G(-2+s)\left(r^{2}(-2+s)(-1+s)+kl^{2}s^{2}\right)}\quad.\label{eq:topobhC}\end{equation}
If we set $\lambda=1$, it reduces to the expression discussed in
\cite{cai2009topological}. We show the special case of $s=\frac{1}{2}\,(\lambda=1)$
in Figure \ref{fig:TopoC}.

\begin{figure}[H]
\begin{centering}
\subfloat[$k=1$]{\centering{}\includegraphics[scale=0.66]{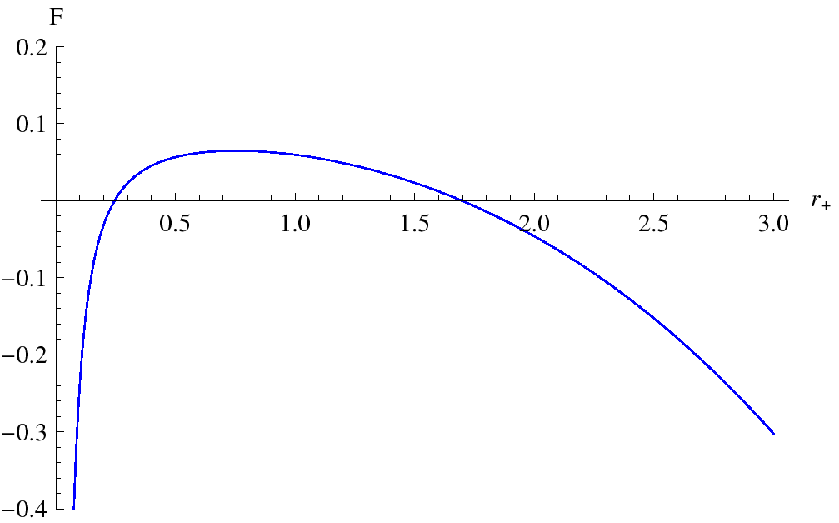}}\subfloat[$k=0$]{\centering{}\includegraphics[scale=0.66]{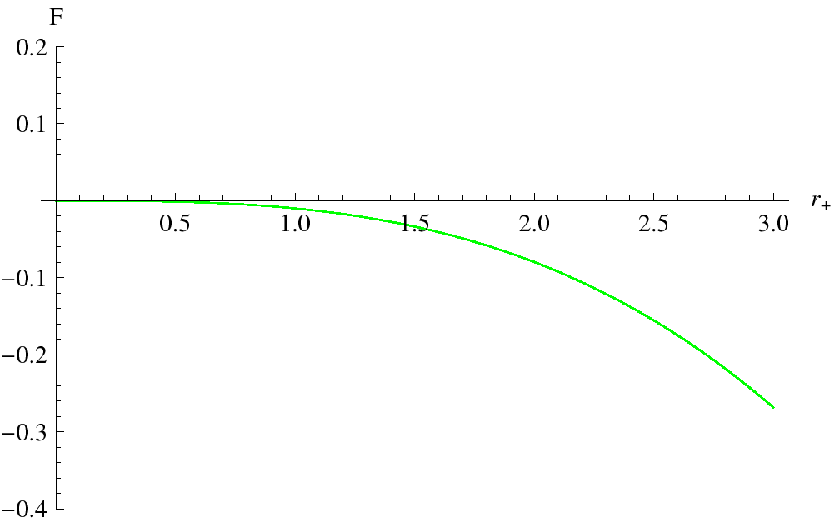}}
\par\end{centering}

\centering{}\subfloat[$k=-1$\label{fig:topoFkm1}]{\centering{}\includegraphics[scale=0.7]{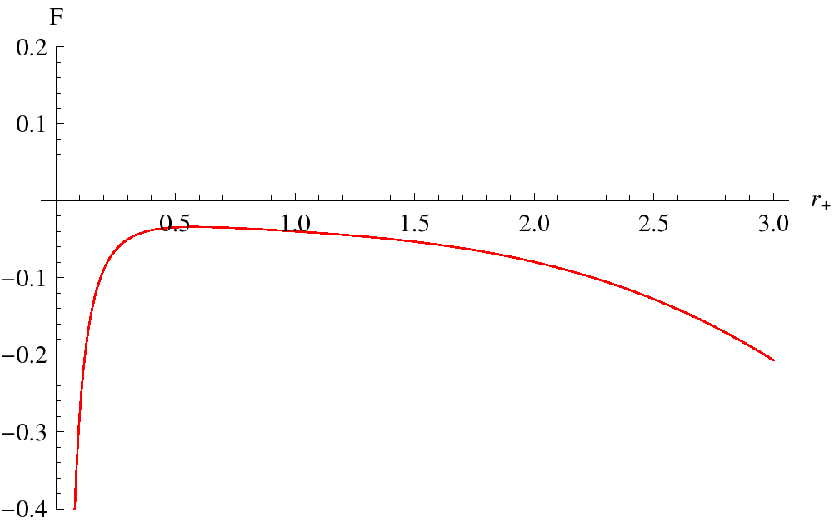}}\caption{Free energy for topological black holes \emph{vs}. horizon radii $r_{+}$,
with $c=l=G=\gamma=\Omega_{k}=1$ and $s=\frac{1}{2}$.\label{fig:TopoF}}

\end{figure}

From Figure \ref{fig:TopoCkm1}, we see that there is a divergent
point of the heat capacity for $k=-1$. The divergent point, $r_{+}=\frac{s}{\sqrt{s^{2}-3s+2}}$,
coincides with the value of $r_{+}$ for minimum temperature. Across
this point, the local stability of black hole changes. In order to
study the phase structure and global stability, we must observe the
free energy, it can be obtained as follows:\begin{equation}
F=M-TS=-\frac{c^{3}(1+s)\gamma\left(r^{-2s}\left(kl^{2}+r^{2}\right)^{2}+r^{-2s}\left(r^{2}(-2+s)+kl^{2}s\right)\left(r^{2}+kl^{2}\ln\left(\frac{r^{2}}{l^{2}}\right)\right)\right)\Omega_{k}}{16Gl^{2}\pi(-2+s)}\quad.\label{eq:topoF}\end{equation}
To see its properties, we plot the free energy for $s=\frac{1}{2}$
($\lambda=1$) in Figure \ref{fig:TopoF}. The maximum point of the
free energy for $k=-1$ in Figure \ref{fig:topoFkm1} is $r_{+}=\frac{s}{\sqrt{2-3s+s^{2}}}$,
which coincides with with the point for the minimum temperature and
divergent heat capacity. %
\begin{figure}[H]
\begin{centering}
\includegraphics{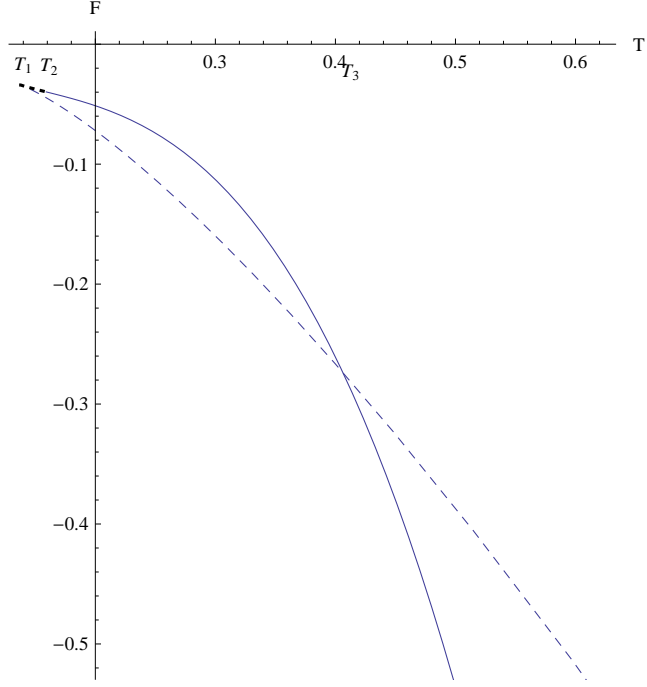}
\par\end{centering}

\caption{Free energy for topological black holes with $k=-1$ \emph{vs}. temperature,
with $c=l=G=\gamma=\Omega_{k}=1$ and $s=\frac{1}{2}$. The thick
dotted curve is for the middle branch, while the dashed and solid
curves are associated with the small and large branches, respectively.
Note that the origin coordinates are $(0.2,\,0)$. \label{fig:TopoF to Temp}}

\end{figure}

The free energy as a function of temperature for $k=-1$ is showed
in Figure \ref{fig:TopoF to Temp}. There is no black holes below
a certain temperature $T_{1}$ ($T_{1}=\frac{\sqrt{3}}{4\pi}$ in
the plot). Above $T_{1}$, two black holes can exist and the small
branch is the more stable than the middle branch. As the temperature
increasing, at temperature $T_{2}$ ($T_{2}=\frac{1}{2\pi}$ in the
plot), the large branch take the place of middle branch, but the free
energy of the small branch is still more negative than that of the
large branch. As the temperature increases further, the small and
large branches have the same free energy at $T_{3}$ ($T_{3}$ is
about $0.406$ in the plot), above this temperature, the large branch
takes over the physics. This signifies a phase transition goes from
small to large $r_{+}$ black holes. At $T_{3}$, the free energy
is continuous, but its first derivative is discontinuous. This is
a typical first order phase transition .

The above results can be interpreted as follow. For our specific choice
$s=\frac{1}{2}$, the thermodynamic properties of the topological
black holes in H-L gravity with different horizon topologies are very
different from each other. For $k=1$, the free energy decreases steadily
in the physical region. Any non-extremal black holes have positive
heat capacity, and the black holes are stable. Thermodynamical behavior
of the $k=0$ case is just like the $k=1$ case. The story is very
different for $k=-1$ case. It can be inform from Figure \ref{fig:TopoCkm1}
that there are three branches of black holes. The middle branch is
unstable while the small and large branches are stable. The branch
with the smallest radii is new in H-L gravity, and does not appear
in the Schwarzschild AdS black hole. Also notice that the middle
branch has the highest free energy, and is relatively unstable.

We have found interesting phase structures in the topological black
holes in H-L gravity, some of the properties are never observed in
Einstein gravity. The H-L gravity is regarded as a UV complete candidate
for general relativity, and the ultraviolet behavior of spacetime
changed dramatically, maybe, this is why there is a branch of stable
black hole with a small radii appear in the $k=-1$ case. Pursuing
its deep reason goes beyond the scope of this paper, and we will leave
settling of this interesting issue to a future date. For general dynamical
coupling constant $\lambda$, the heat capacity for three cases with
different horizon topologies can all have positive and negative values,
which means that there always exist locally thermodynamically stable
phases and unstable phases in suitable parameter regimes. This indicate
phase transitions at certain threshold points. Relevant results can
be found in \cite{cai2009thermodynamics}.

In \cite{biswas2010geometry}, the authors obtained the Ruppeiner
metric for this kind of black holes by taking $\lambda$ and $M$
as variables. Weinhold metric was also calculated there. However,
in Ho\v{r}ava-Lifshitz gravity, $\lambda$ represents a dynamical
coupling constant, susceptible to quantum corrections, it is not a
conserved charge, a convincing reason is needed for why it can be
taken as a variable in constructing thermodynamic metrics. Taking
this argument into account, one see that the thermodynamical metric
of the uncharged black hole is one dimensional, so we do not study
the various properties of the scalar curvatures.

\section{Charged black hole in Ho\v{r}ava-Lifshitz gravity\label{sec:Charged-black-hole}}

Phase transitions in charged black holes such as Reissner-Nordstr\"om
black holes have been studied in \cite{davies1977thermodynamic,davies1989thermodynamic,chamblin1999charged,chamblin1999holography,Carlip:2003ne}.In
this section we extend the discussion of phase transitions to charged
black holes in H-L gravity. For simplicity and clearness we set $\lambda=1$.
The charged topological black hole solution in H-L gravity for $\lambda=1$
has been given in \cite{cai2009topological}, the metric is

\begin{eqnarray}
ds^{2} & = & -\tilde{N}(r)^{2}f(r)dt^{2}+\frac{dr^{2}}{f(r)}+r^{2}d\Omega_{k}^{2}\quad,\\
f(r) & = & k+x^{2}-\sqrt{c_{0}x-\frac{q^{2}}{2}}\,,\qquad x=\sqrt{-\Lambda}\, r\quad,\end{eqnarray}
where $\tilde{N}=N_{0}$ could be set to one. The event horizon radius
is determined by $x_{+}$ (note $x_{+}$ is not the horizon radius),
which is the largest positive root of $f(x)=0$. Denoting $l^{2}=-\frac{1}{\Lambda}$
as well, the thermodynamical quantities computed in \cite{cai2009topological}
are\begin{eqnarray}
T & = & \frac{6x_{+}^{4}+4kx_{+}^{2}-2k^{2}-q^{2}}{16kl^{2}\pi x_{+}+16l^{2}\pi x_{+}^{3}}\quad,\\
S & = & \frac{\pi\kappa^{2}\mu^{2}\Omega_{k}}{4}\left(x_{+}^{2}+2k\,\ln\left(x_{+}\right)\right)+S_{0\quad,}\\
\Phi & = & \frac{q}{x_{+}}+\Phi_{0}\quad,\\
Q & = & \frac{\kappa^{2}\mu^{2}\Omega_{k}}{16l^{2}}q\quad,\\
M & = & \frac{\kappa^{2}\mu^{2}\Omega_{k}}{16l^{2}}c_{0}\quad,\end{eqnarray}
where $c_{0}=\frac{2k^{2}+q^{2}+4kx_{+}^{2}+2x_{+}^{4}}{2x_{+}}$.
As we can see, the mass of the charged topological black hole in H-L
gravity is also not increasing monotonously with $x_{+}$. It is more
complicate than the uncharged one, since the charge parameter $q$
comes into play. The charge parameter $q$ does not appear explicitly
in the expression of entropy $S$. It is consistent with the fact
that black hole entropy is a function of horizon geometry. It is easy
to verify that the first law of thermodynamic $dM=TdS+\Phi dQ$ is
satisfied.

\begin{figure}[H]
\begin{centering}
\subfloat[$k=1$\label{fig:chargedTk1}]{\includegraphics[scale=0.66]{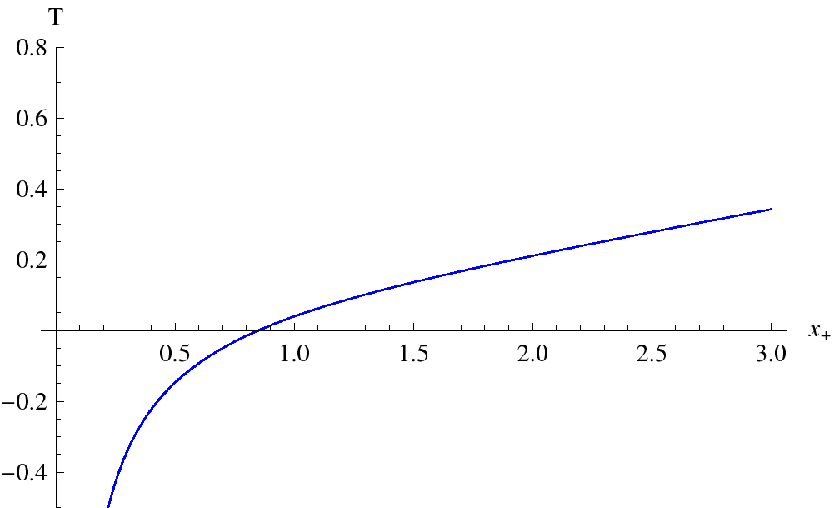}}\subfloat[$k=0$\label{fig:chargedTk0}]{\includegraphics[scale=0.66]{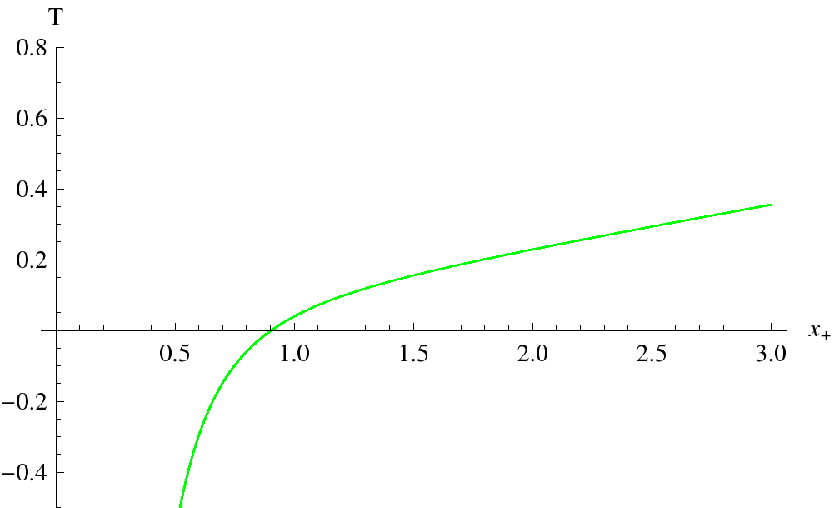}}
\par\end{centering}

\centering{}\subfloat[$k=-1$\label{fig:chargedTkm1}]{\includegraphics[scale=0.7]{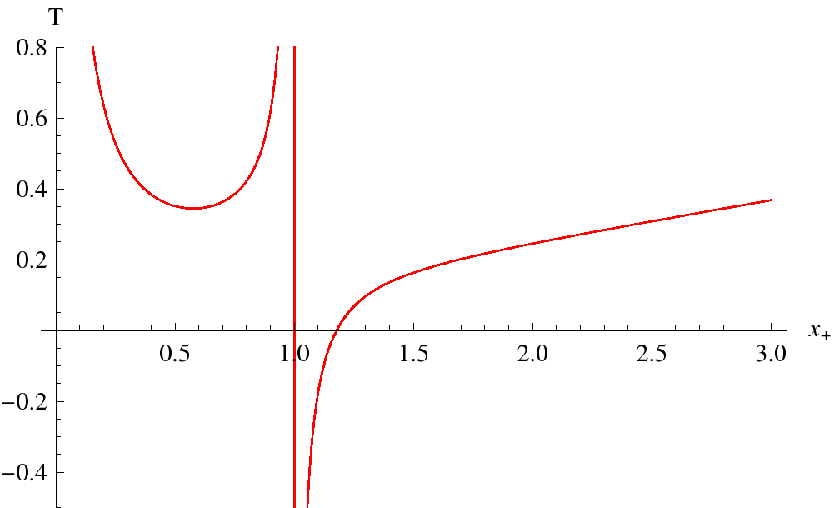}}\caption{Temperature of charged topological black holes \emph{vs. $x_{+}$},
the values $c=l=G=\gamma=\Omega_{k}=1$ and $q=2$ have been used
here. Each of this three graphs have unphysical regions with negative
Hawking temperature. The unusual things appeared in the last graph
with $k=-1$. See text for the discussion about this. \label{fig:ChargedT}}

\end{figure}

\begin{figure}[H]
\centering{}\includegraphics[scale=0.8]{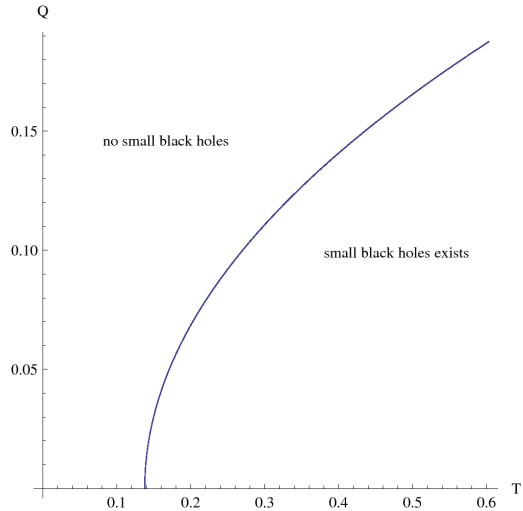}\caption{Phase structure of {}``small'' ($0<x_{+}<1$) topological black
holes for $k=-1$.\label{fig:Chargedphase}}

\end{figure}

When $q^{2}=2(-k^{2}+2kx_{+}^{2}+3x_{+}^{4})$, or $x_{+}=\sqrt{-\frac{k}{3}+\frac{\sqrt{8k^{2}+3q^{2}}}{3\sqrt{2}}}$
, Hawking temperature $T=0$, and the black hole becomes an extremal
one. In order for this black hole to be non-extremal we must assume\[
q^{2}<2(-k^{2}+2kx_{+}^{2}+3x_{+}^{4})\,.\]
We exhibit the temperature \emph{vs. $x_{+}$ }curve in Figure \ref{fig:ChargedT}
, with parameters $c=l=G=\gamma=\Omega_{k}=1$ and $q=2$ (Parameters
are chose in the same way in all the graphical analyzes in this section,
but the general properties of the physical quantities are not changed
 if we choose other parameters). We can see that each of the three
graphs have negative values of Hawking temperature at some ranges
of $x_{+}$, i.e. $q^{2}>2(-k^{2}+2kx_{+}^{2}+3x_{+}^{4})$, which
correspond to unphysical regions and have no physical significance.
There is an infinite discontinuous point $x_{+}=x_{c}=1$ in the case
of $k=-1$ (Figure \ref{fig:chargedTkm1}). When $x_{+}>x_{c}$, the
temperature behaves just like the other two temperature curves in
Figures \ref{fig:chargedTk1} and \ref{fig:chargedTk0}, gradually
increasing with $x_{+}$. Another interesting property is that in
the case of $k=-1$, for {}``small'' black holes (with radius $0<x_{+}<1$),
the temperature has a minimum value at the point $x_{+}=\frac{1}{\sqrt{3}}$.
Remarkably, this point does not depend on the charge parameter $q$.
We plot the phase structure of small black holes in Figure \ref{fig:Chargedphase}.

As emphasized in Section \ref{sec:Topological-black-hole}, the temperature
\emph{vs. $x_{+}$ }curve cannot reveal all the information of the
phase structure, but it tells us the {}``small'' black holes only
exists above a minimum temperature, as shown in Figure \ref{fig:Chargedphase}.

The charged topological black holes in H-L gravity have two conserved
quantities $Q$ and $M$ parameterized by $q$ and $x_{+}$, just
like the RN AdS black holes. The RN AdS black holes phase transitions
were studied in \cite{chamblin1999charged,chamblin1999holography}.
Two complementary thermodynamic ensembles were studied there, which
are the grand canonical ensemble with fixed background potential and
the canonical ensemble with fixed charge. In the canonical ensemble,
if the fixed charge is below a critical value, with the temperature
increases, small black holes are tend to become large black holes
with a jump in the entropy. This is a first order phase transition
with a release of {}``latent heat''. Above the critical value of
the charge, there is no phase transition. This is very similar to
to the van der Waals-Maxwell liquid-gas model. This phase transition
is characterized by the divergence of the specific heat with fixed
charge $C_{Q}$. Recently, a new type of phase transition for the
RN AdS black hole associated with the divergence of the specific heat
with fixed potential $C_{\Phi}$ was suggested in \cite{Banerjee:2010da}.
The authors argued that study of the Ehrenfest's equations and the
thermodynamical metrics indicate the existence of a new glassy type
transition.

Now, let's proceed to do the stability anolysis for charged topological black
holes in H-L gravity. Thermodynamic stability could be studied in
many different ways, depending on which thermodynamic function we
choose to use. Different choices will reveal thermodynamic stability
in different ensembles. A physical transparent way is to examine the
sign of specific heats and other capacities. For charged topological
black holes in H-L gravity in hand, four kinds of capacities can be
constructed. They are\begin{equation}
C_{Q}\equiv T\frac{\partial S}{\partial T}\biggl|_{Q}\quad,\qquad C_{\Phi}\equiv T\frac{\partial S}{\partial T}\biggl|_{\Phi}\quad,\qquad\tilde{C_{T}}\equiv\frac{\partial Q}{\partial\Phi}\biggl|_{T}\quad,\qquad\tilde{C_{S}}\equiv\frac{\partial Q}{\partial\Phi}\biggl|_{S}\quad.\label{eq:defination of four C}\end{equation}
The first two, i.e., the specific heats at constant electric charge
or potential, determine the thermal stability of the black holes.
They are positive (negative) if the black hole is thermodynamic stable
(unstable) to a thermal fluctuation. The last two quantities are charge
capacitances at fixed temperature or entropy. They are negative (positive)
if the black hole is electrically unstable (stable) to electrical
fluctuation.

Following the standard thermodynamic definition of specific heat,
we calculate the specific heat of the topological charged black hole
in H-L gravity for constant charge $Q$, \begin{equation}
C_{Q}\equiv T\frac{\partial S}{\partial T}\biggl|_{Q}=\frac{\partial M}{\partial T}\biggl|_{Q}=-\frac{\pi\left(k+x_{+}^{2}\right)^{2}\left(q^{2}+2\left(k-3x_{+}^{2}\right)\left(k+x_{+}^{2}\right)\right)\kappa^{2}\mu^{2}\Omega_{k}}{2\zeta_{1}\left(q^{2}+2\left(k+x_{+}^{2}\right)^{2}\right)}\quad,\label{eq:ChargedCQ}\end{equation}
where\begin{equation}
\zeta_{1}=\left(k+3x_{+}^{2}\right)\quad.\end{equation}
Note that the expressions of $C_{Q}$ and $T$ share the same factor
$q^{2}+2\left(k-3x_{+}^{2}\right)\left(k+x_{+}^{2}\right)$ in their
numerator, and the non-extremal condition will make $C_{Q}$ positive
definite in the physical parameters region when $k=0,\,1$, which
means they are thermodynamic stable locally.

The specific heat for constant potential $\Phi$ is given by\begin{eqnarray}
C_{\Phi} & \equiv & T\frac{\partial S}{\partial T}\biggl|_{\Phi}=-\frac{\pi\left(k+x_{+}^{2}\right)^{2}\left(-q^{2}+2\left(k-3x_{+}^{2}\right)\left(k+x_{+}^{2}\right)\right)\kappa^{2}\mu^{2}\Omega_{k}}{2\zeta_{2}}\quad,\end{eqnarray}
where \begin{eqnarray}
\zeta_{2} & = & \left(2k^{3}-kq^{2}+10k^{2}x_{+}^{2}+q^{2}x_{+}^{2}+14kx_{+}^{4}+6x_{+}^{6}\right)\quad.\end{eqnarray}
We can also calculate the charge capacitances at fixed temperature
or entropy\begin{eqnarray}
\tilde{C_{T}} & \equiv & \frac{\partial Q}{\partial\Phi}\biggl|_{T}=\frac{1}{16l^{2}}\left(x_{+}+\frac{2q^{2}x_{+}\left(k+x_{+}^{2}\right)}{\zeta_{2}}\right)\kappa^{2}\mu^{2}\Omega_{k}\quad,\\
\tilde{C_{S}} & \equiv & \frac{\partial Q}{\partial\Phi}\biggl|_{S}=\frac{1}{16l^{2}}x_{+}\kappa^{2}\mu^{2}\Omega_{k}\quad.\label{eq:capacity S}\end{eqnarray}

\begin{figure}[H]
\begin{centering}
\subfloat[$k=1$]{\centering{}\includegraphics[scale=0.66]{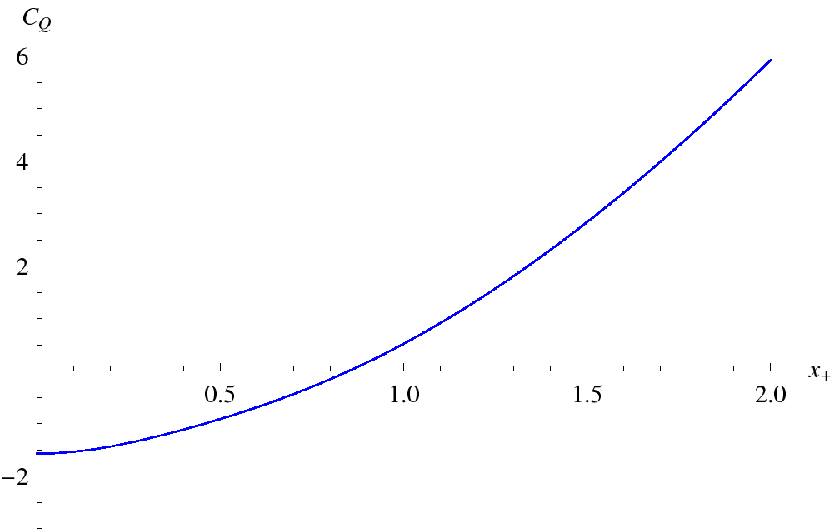}}\subfloat[$k=0$]{\centering{}\includegraphics[scale=0.66]{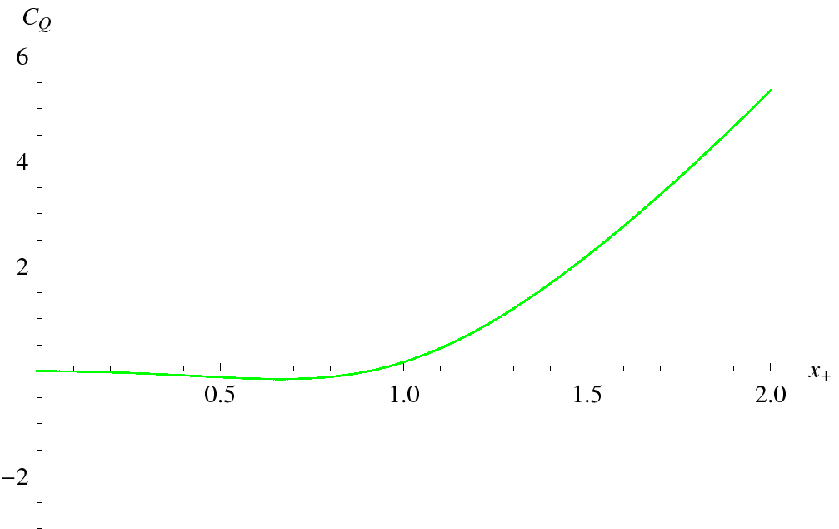}}
\par\end{centering}

\centering{}\subfloat[$k=-1$\label{fig:ChargedCQ km1}]{\centering{}\includegraphics[scale=0.7]{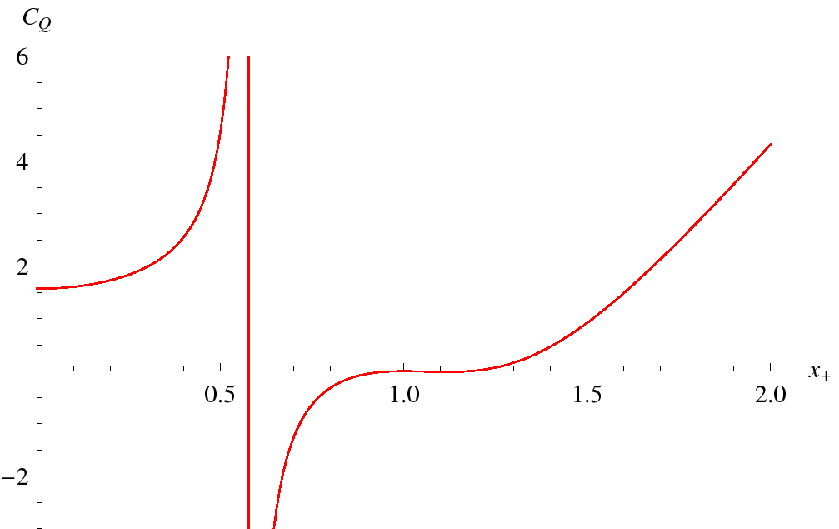}}\caption{Specific heat for constant charge $Q$ \emph{vs. $x_{+}$}, with $c=l=G=\gamma=\Omega_{k}=1$
and $q=2$. $C_{Q}$ is positive definite in the physical region of
parameters when $k=0,\,1$, it can change it's sign at a parameter
$q$ independent threshold point $x_{+}=\frac{1}{\sqrt{3}}$ in the
case of $k=-1$. \label{fig:ChargedCQ}}

\end{figure}
We plot $C_{Q}$ in Figure \ref{fig:ChargedCQ}, with parameters $c=l=G=\gamma=\Omega_{k}=1$
and $q=2$. As we can see from Figure \ref{fig:ChargedCQ km1}, the
specific heat passes from positive to negative values through an infinite
discontinuity, and then passes across the horizontal ordinate again
continuously (a small range of $x_{+}$ bigger than $x_{+}=1$ is
unphysical since the Hawking temperature is negative). The first threshold
point is associated with a local phase transition, and the latter
happens at Hawking temperature $T=0$ where the black hole becomes
an extremal one. The latter threshold point also occur in the cases
of $k=1$ and $k=0$. These points are not associated with phase transitions.
However, they define the unphysical range of $x_{+}$, where the temperature
acquires a negative value and the curve does not have any physical
significance. The strange behavior of the specific heat for $k=-1$
should be emphasized. $C_{Q}$ has a divergent point in the case of
$k=-1$ at $\zeta_{1}=0$, i.e., $x_{+}=\frac{1}{\sqrt{3}}$, which
is consistent with the minimum of temperature. This indicates a local
phase transition.

For brevity, we are not going to show all the other three $C$'s in
graphs. We can see that $C_{Q}$, $C_{\Phi}$ and $\tilde{C}_{T}$
all can have both positive and negative values in certain parameter
regions, they will change signs in threshold points. The threshold
points are at certain value of $x_{+}$, $p$ and $k$ which satisfy
the condition \begin{eqnarray*}
\zeta_{1}=0 & \quad or\quad & \zeta_{2}=0\quad.\end{eqnarray*}
These threshold points signal the change of local stabilities and
hence are typically associated with certain local phase transitions.
$\tilde{C_{S}}$ is positive definite, so it may not be relevant to
phase transition.

\begin{figure}[H]
\begin{centering}
\subfloat[$k=1$]{\centering{}\includegraphics[scale=0.66]{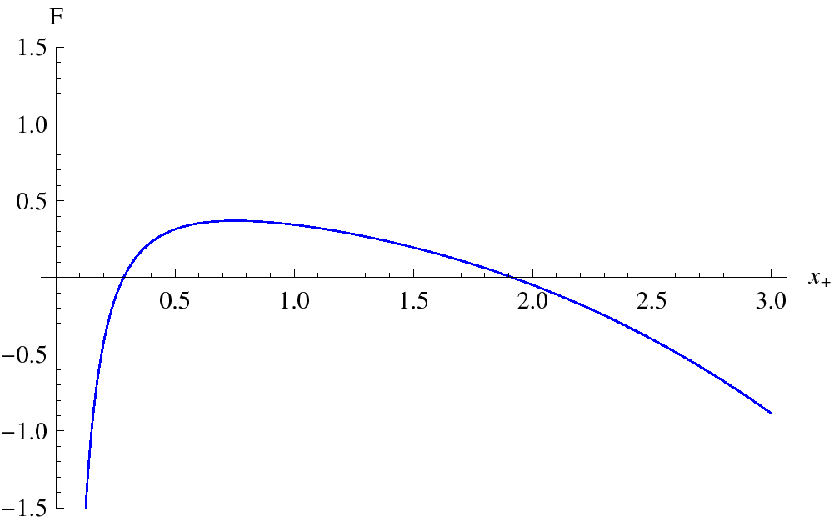}}\subfloat[$k=0$]{\centering{}\includegraphics[scale=0.66]{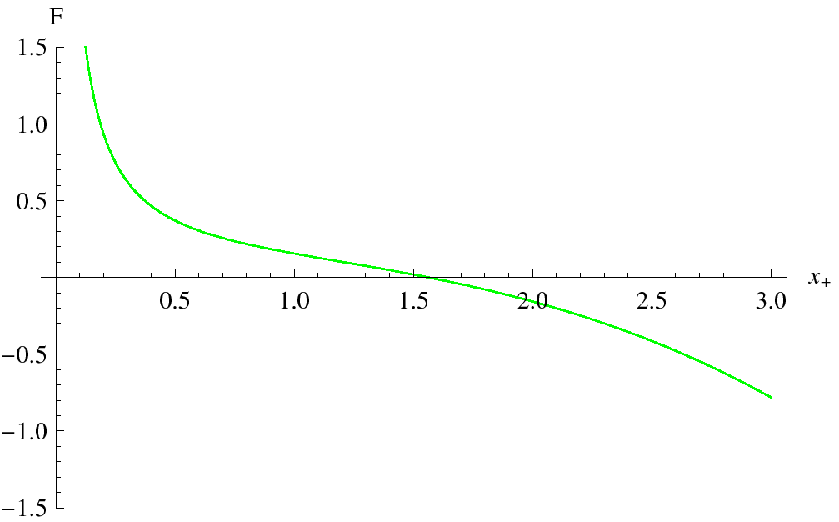}}
\par\end{centering}

\centering{}\subfloat[$k=-1$\label{fig:ChargedFkm1}]{\centering{}\includegraphics[scale=0.7]{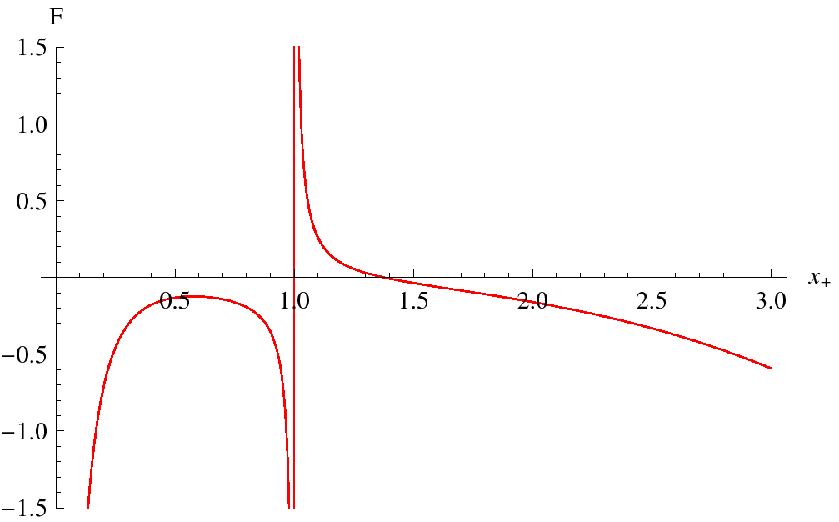}}\caption{Free energy for constant charge $Q$ \emph{vs. $x_{+}$}, the values
$c=l=G=\gamma=\Omega_{k}=1$ and $q=2$ have been used here, see text
for discussion. \label{fig:ChargedF}}

\end{figure}

\begin{figure}[H]
\begin{centering}
\includegraphics{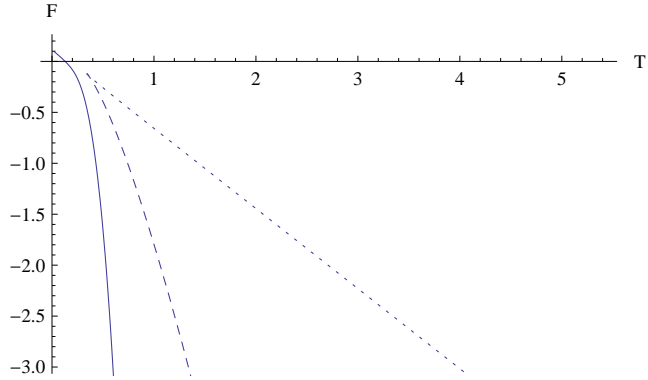}
\par\end{centering}

\caption{Free energy for charged topological black holes with $k=-1$ \emph{vs}.
temperature, with $c=l=G=\gamma=\Omega_{k}=1$ and $q=2$. The dotted
curve is for the middle branch, while the dashed and solid curves
are associated with the small and large branches, respectively. \label{fig:chargeF to Temp}}

\end{figure}

The free energy can reveal global stability properties of black holes.
It is straightforward to calculated as \begin{eqnarray}
F & = & M-TS\nonumber \\
 & = & \frac{\kappa^{2}\mu^{2}\Omega_{k}\left(4k^{3}+14k^{2}x_{+}^{2}+3q^{2}x_{+}^{2}-2x_{+}^{6}+2k\,\ln\left(x_{+}\right)\left(2k^{2}+q^{2}-4kx_{+}^{2}-6x_{+}^{4}\right)+2k\left(q^{2}+4x_{+}^{4}\right)\right)}{64l^{2}x_{+}\left(k+x_{+}^{2}\right)}\,.\end{eqnarray}
We plot the free energy as a function of $x_{+}$ in Figure \ref{fig:ChargedF},
with $c=l=G=\gamma=\Omega_{k}=1$ and $q=2$. For $k=0,\,1$, the
free energy decrease steadily in the physical region. Any non-extremal
black holes have positive heat capacity, and the black holes are stable.
Now, let's focus on Figure \ref{fig:ChargedFkm1}. For ``small''
black holes in the case of $k=-1$, the free energy has a maximum
at $x_{+}=\frac{1}{\sqrt{3}}$, which is also the point of minimum
temperature and divergent heat capacity $C_{Q}$. As we can see from
Figure \ref{fig:ChargedCQ km1}, there are three branches of black
holes. In the region $\frac{1}{\sqrt{3}}<x_{+}<1$, the middle branch
non-extremal black holes have negative value of $C_{Q}$, and they
are local unstable, while the small and large branches are local stable.
One may suspect that there is a global phase transition between the
small and large $x_{+}$ black holes. This does not take place since
the large branch of black holes persist to dominate the thermodynamics
at all temperatures. As we can see from Figure \ref{fig:chargeF to Temp},
at all temperatures, the large branch (solid curve in the graph) has
the most negative value of free energy, and global phase transition
can not comes up. %

An alternative way to study thermodynamics is using thermodynamical
metrics, which have been introduced in Section \ref{sec:introduction}.
Next, we are going to study the properties of the corresponding state
space geometry. The Ruppeiner and Weinhold metrics, and the thermodynamical
metric derived from the free energy $F=M-T\, S$ are defined by\begin{equation}
\vec{M}=\left\{ M^{i}\right\} \equiv\left\{ M,N^{a}\right\} \quad,\qquad g_{ij}^{\left(S\right)}\equiv\frac{\partial^{2}S\left(M,N^{a}\right)}{\partial M^{i}\partial M^{j}}\quad,\end{equation}
\begin{equation}
\vec{S}=\left\{ S^{i}\right\} \equiv\left\{ S,N^{a}\right\} \quad,\qquad g_{ij}^{\left(M\right)}\equiv\frac{\partial^{2}M\left(S,N^{a}\right)}{\partial S^{i}\partial S^{j}}\quad,\end{equation}
\begin{equation}
\vec{T}=\left\{ T^{i}\right\} \equiv\left\{ T,N^{a}\right\} \quad,\qquad g_{ij}^{\left(F\right)}\equiv\frac{\partial^{2}F\left(T,N^{a}\right)}{\partial T^{i}\partial T^{j}}\quad.\end{equation}

The Ricci scalars of the above thermodynamical metrics can be calculated
parallel to \cite{liu2010thermodynamical}. The thermodynamical curvatures
are too complicated to present, so we only give the denominators of
the Ricci scalars here, which are enough for us to analyze their singular
behaviors. The denominators of the Ricci scalars, denoted by $D(R)$,
are given by

\begin{eqnarray}
D(R^{\left(S\right)}) & = & \zeta_{2}\pi\kappa^{2}\mu^{2}\Omega_{k}\left(k+x^{2}\right)\left(8+q^{2}-8x_{+}^{2}-6x_{+}^{4}\right)^{4}\left(2+q^{2}-4x_{+}^{2}-6x_{+}^{4}\right)^{4}\left(2+3x_{+}^{2}+x_{+}^{4}\right)^{2}\nonumber \\
 &  & \left(q^{2}\left(-1+x_{+}^{2}\right)+2\left(1+x_{+}^{2}\right)^{2}\left(1+3x_{+}^{2}\right)\right)^{2}\left(q^{2}\left(-2+x_{+}^{2}\right)+2\left(2+x_{+}^{2}\right)^{2}\left(2+3x_{+}^{2}\right)\right)^{2}\quad,\\
D(R^{\left(M\right)}) & = & \frac{\kappa^{2}\mu^{2}\Omega_{k}}{l^{2}}\zeta_{2}\left(q^{2}\left(-1+x_{+}^{2}\right)+2\left(1+x_{+}^{2}\right)^{2}\left(1+3x_{+}^{2}\right)\right)^{2}\nonumber \\
 &  & \left(q^{2}\left(-2+x_{+}^{2}\right)+2\left(2+x_{+}^{2}\right)^{2}\left(2+3x_{+}^{2}\right)\right)^{2}\quad,\\
D(R^{\left(F\right)}) & = & \frac{\kappa^{2}\mu^{2}\Omega_{k}}{l^{2}}\left(q^{2}+2\left(1+x_{+}^{2}\right)^{2}\right)^{4}\left(q^{2}+2\left(2+x_{+}^{2}\right)^{2}\right)^{4}\quad.\end{eqnarray}

We can see that the curvature singularity of these metrics are related
to the local phase transition associated with the vanishing of $\zeta_{2}$,
not $\zeta_{1}$. In the cases discussed in \cite{liu2010thermodynamical},
the curvature singularity of the Ricci scalar for the free-energy
metric $R^{\left(F\right)}$ is associated with the divergence of
the specific heat at $\zeta_{1}=0$. However, this does not happen
in the case of topological charged black hole in H-L gravity. The
denominator of $C_{Q}$ and $R^{\left(F\right)}$ indeed share the
same factor \begin{equation}
\zeta_{3}=q^{2}+2\left(k+x_{+}^{2}\right)^{2}\quad.\end{equation}
But $\zeta_{3}$ does not vanish unless we set $q=0$, which reduce
to the uncharged case discussed in Sec.(\ref{sec:Topological-black-hole}).
So we have to declare that one exception to the framework in \cite{liu2010thermodynamical}
has been found. The specific heat $C_{Q}$ of charged black hole in
H-L gravity with $k=-1$ is divergent at $x_{+}=\frac{1}{\sqrt{3}}$,
but none of the three Ricci scalars have singularity at this point.
The violation of the correspondence only exists when $k=-1$, this
happens because the topology of its event horizon is non-spherical.
This is a very interesting feature of H-L gravity.

In closing this section, we would like to stress the unusual features
we have found here. The phase structures is vivid in the charged topological
black holes in H-L gravity, which are very different from the RN AdS
black holes in Einstein gravity. For the charged black holes with
sphere topology in H-L gravity, there is no phase transition in the
canonical ensemble with fixed charge, since the specific heat with
fixed charge $C_{Q}$ is positive definite, and all the non-extremal
black holes can local stable exist at all temperatures. Whilst, this
is not the performance of RN AdS black holes. Another difference is
that the thermodynamical metrics can not reveal all of the local phase
transition signals in H-L gravity, such as the charged black holes
in H-L gravity with non-spherical event horizon. One should also note
that this is a local phase transition, global phase transitions are
not take place in the canonical ensemble with fixed charge for the
charged topological black holes in H-L gravity. As we have declared
in the end of Section \ref{sec:Topological-black-hole}, the ultraviolet
behavior of spacetime in H-L gravity deserves as the reason for all
of those strange properties in black hole thermodynamics.

\section{Kihagias-Sfetosos black hole\label{sec:KS-black-hole}}

The action of the IR modified Ho\v{r}ava-Lifshitz gravity is obtained
by adding a term $\mu^{4}R^{\left(3\right)}$ to the original Ho\v{r}ava-Lifshitz
action\cite{kehagias2009black}, in order to get a Minkowski vacuum
in IR. The Kihagias-Sfetosos (KS) black hole is a solution of the
IR modified (or deformed) Ho\v{r}ava-Lifshitz gravity, which was
originally obtained in \cite{kehagias2009black}. The thermodynamical
properties of this kind of black holes has been studied in \cite{myung2009thermodynamics,myung2010entropy,castillo906entropy,wei2010thermodynamic}.
In this section, we shall study phase transitions in KS black holes.
Stability analysis and state space geometry will both be used. The
solution of KS black hole is defined by the line element\begin{equation}
ds^{2}=-N^{2}(r)dt^{2}+\frac{1}{f(r)}dr^{2}+r^{2}\left(d\theta^{2}+sin^{2}\theta d\phi^{2}\right)\quad,\end{equation}
where\begin{equation}
N^{2}=f=1+\omega r^{2}-\sqrt{r(\omega^{2}r^{3}+4\omega m)}\quad,\label{eq:KS solution}\end{equation}
and $\omega=\frac{16\mu^{2}}{\kappa^{2}}$ is a combined parameter
of the original parameters $\kappa$ and $\mu$ in H-L gravity action.
The quantity $\sqrt{\frac{1}{2\omega}}$ behaves as a charge-like
parameter, we denote $P=\sqrt{\frac{1}{2\omega}}$ and consider it
as a new parameter in the black hole thermodynamics. Now, we can rewrite
(\ref{eq:KS solution}) as \begin{equation}
N^{2}(r,P)=f(r,P)=1+\frac{r^{2}}{2P^{2}}-\sqrt{\frac{r^{4}}{4P^{4}}+\frac{2mr}{P^{2}}}\quad.\end{equation}
The horizons of KS black hole is determined by $f(r,P)=0$, which
gives\begin{equation}
r_{\pm}=m\pm\sqrt{m^{2}-P^{2}}\quad.\end{equation}

The KS black hole is asymptotically flat, and it's thermodynamic properties
will change dramatically from the charged topological black holes
in H-L gravity. The mass parameter $m$ and temperature of the KS
black hole can be expressed respectively as \begin{eqnarray}
m & = & \frac{r_{+}}{2}+\frac{P^{2}}{2r_{+}}=\frac{r_{+}+r_{-}}{2}\quad,\label{eq:KSm}\\
T & = & \frac{f^{\prime}(r)}{4\pi}\biggl|_{r=r_{+}}=\frac{r_{+}^{2}-P^{2}}{4\pi r_{+}\left(2P^{2}+r_{+}^{2}\right)}=\frac{r_{+}-r_{-}}{4\pi r_{+}\left(r_{+}+2r_{-}\right)}\quad.\label{eq:KSTem}\end{eqnarray}
The KS black hole becomes an extreme one when $r_{+}=r_{-}$, i.e.,
$m=r_{+}=\left|P\right|$. If one regard $m$ as the mass of KS black
hole, using the first law $dm=T\, dS$, the entropy can be obtained
via integration \begin{equation}
S=\pi r_{+}^{2}+2\pi P^{2}\ln r_{+}^{2}+S_{0}\quad.\label{eq:KSentropyLog}\end{equation}
This is an entropy with logarithmic term. There are many discussions
for Eq.(\ref{eq:KSentropyLog}). In \cite{wei2010thermodynamic},
the authors regard the existence of logarithmic term in the entropy
as a unique feature of the Ho\v{r}ava-Lifshitz gravity. In \cite{myung2010entropy},
it was suggested that this logarithmic term can be interpreted as
the GUP-inspired black hole entropy, and a duality between the entropy
of the KS black hole and the GUP-inspired Schwarzschild black hole
has been shown. On the other hand, one may treat $m$ as a mass parameter
and use the ADM mass to write its first law of thermodynamics. The
authors of \cite{myung2010adm} and \cite{wang912first} argued that
one should use the area-law entropy and ADM mass, which also satisfy
the first law of thermodynamics.  Here, we examine the thermodynamical
properties of KS black holes by using the first idea, and the latter
idea will be adopt in the appendix for completeness.

If we take $P$ as the KS black hole charge, the potential $\Phi$
corresponds to  $P$ is \begin{equation}
\Phi=\frac{P\left(2P^{2}+r_{+}^{2}+(P-r_{+})(P+r_{+})\ln\left[r_{+}^{2}\right]\right)}{2P^{2}r_{+}+r_{+}^{3}}\quad.\end{equation}
Those thermodynamical quantities also satisfy the first law of thermodynamic
$dm=TdS+\Phi dP$.

\begin{figure}[H]
\centering{}\includegraphics{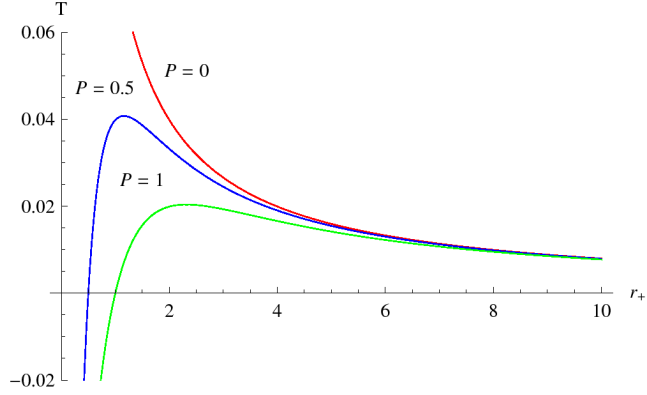}\caption{Temperature of KS black holes.\label{fig:KSlogT}}

\end{figure}

\begin{figure}[H]
\centering{}\includegraphics[scale=0.8]{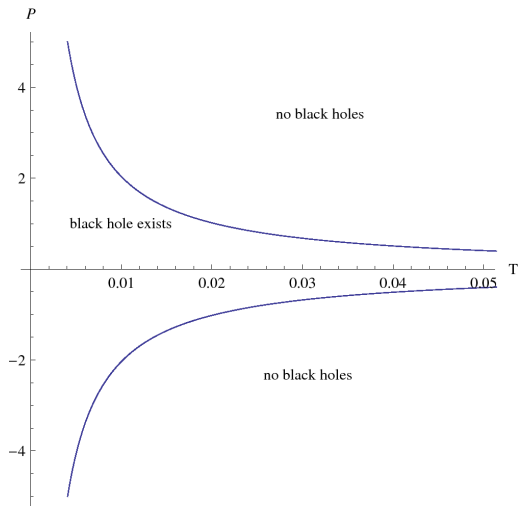}\caption{Phase structure of KS black holes. KS black holes can exist at all
temperatures.\label{fig:KSlogPhase}}

\end{figure}

The Hawking temperature of KS black hole behaves very similar to RN
black hole. As depicted in Figure \ref{fig:KSlogT}, the KS black
hole has a maximum temperature for $P\neq0$, when $2P^{4}+5P^{2}r_{+}^{2}-r_{+}^{4}=0$,
i.e., \begin{equation}
r_{+}=r_{m}=\sqrt{\frac{(5+\sqrt{33})P^{2}}{2}}\label{eq:maximum temperature point ks}\end{equation}
which satisfies $dT/dr_{+}=0$. And we will found that it is just
the threshold point for specific heat changing from negative to positive
values through an infinite discontinuity. Hence, for $P\neq0$, the
point at $r_{m}$ may signal a local phase transition. We plot the
phase structure for KS black holes in Figure \ref{fig:KSlogPhase}.
We can find that there is another similarity between RN black holes
and KS black holes that both of these black holes can exist at all
temperatures.

Taking $P$ and $r_{+}$ as variables, we can calculate the specific
heat for either constant $P$ or constant potential $\Phi$, and charge
capacitances defined in (\ref{eq:defination of four C}). They are
given by\begin{eqnarray}
C_{P} & \equiv & T\frac{\partial S}{\partial T}\biggl|_{P}=\frac{\partial m}{\partial T}\biggl|_{P}=\frac{2\pi\left(-4P^{6}+3P^{2}r_{+}^{4}+r_{+}^{6}\right)}{\xi_{1}}\quad,\label{eq:cp for KS with log}\\
C_{\Phi} & \equiv & T\frac{\partial S}{\partial T}\biggl|_{\Phi}\nonumber \\
 & = & \frac{\left(2\pi\left(2P^{2}+r_{+}^{2}\right)\left(-4P^{6}+12P^{4}r_{+}^{2}+9P^{2}r_{+}^{4}+r_{+}^{6}+\left(2P^{6}+7P^{4}r_{+}^{2}+4P^{2}r_{+}^{4}-r_{+}^{6}\right)\ln\left(r_{+}^{2}\right)\right)\right)}{\left(P^{2}-r_{+}^{2}\right)\xi_{2}}\quad,\\
\tilde{C_{T}} & \equiv & \frac{\partial Q}{\partial\Phi}\biggl|_{T}=\frac{4P^{6}r_{+}+12P^{4}r_{+}^{3}+3P^{2}r_{+}^{5}-r_{+}^{7}}{\left(P^{2}-r_{+}^{2}\right)\xi_{2}}\quad,\\
\tilde{C_{S}} & \equiv & \frac{\partial Q}{\partial\Phi}\biggl|_{S}=\frac{\left(r_{+}\left(2P^{2}+r_{+}^{2}\right)^{3}\right)}{\xi_{3}}\quad,\end{eqnarray}
where\begin{eqnarray}
\xi_{1} & = & 2P^{4}+5P^{2}r_{+}^{2}-r_{+}^{4}\quad,\\
\xi_{2} & = & 4P^{4}+16P^{2}r_{+}^{2}+r_{+}^{4}+\left(2P^{4}+5P^{2}r_{+}^{2}-r_{+}^{4}\right)\ln\left(r_{+}^{2}\right)\quad,\\
\xi_{3} & = & \left(2P^{2}+r_{+}^{2}\right)^{3}+\ln\left(r_{+}^{2}\right)\left(4P^{6}+24P^{4}r_{+}^{2}+9P^{2}r_{+}^{4}-r_{+}^{6}+2\left(2P^{6}+5P^{4}r_{+}^{2}-P^{2}r_{+}^{4}\right)\ln\left(r_{+}^{2}\right)\right)\quad.\end{eqnarray}

\begin{figure}[H]
\centering{}\includegraphics{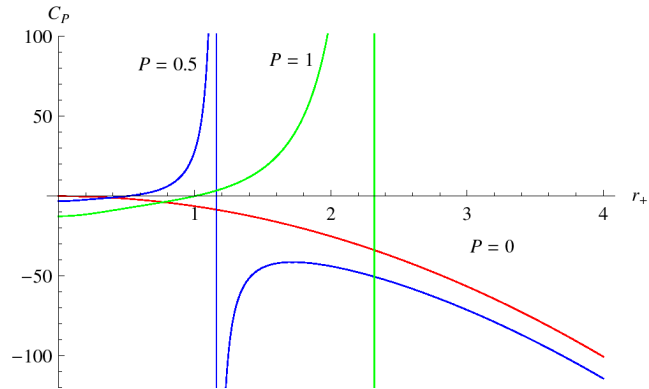}\caption{Heat capacity of KS black holes.\label{fig:KSlogCP}}

\end{figure}

Since $r_{+}^{2}>P^{2}$ for a non-extremal KS black hole, the numerator
of $C_{P}$ is positive definite, and it is divergent only when $\xi_{1}=0$.
This happens at $x_{+}=x_{m}$ which is just the maximum point of
temperature,  as we showed in Eq.(\ref{eq:maximum temperature point ks})
and Figure \ref{fig:KSlogT}. A graph of specific heat $C_{P}$ is
shown in Figure \ref{fig:KSlogCP}.%
 One can see from Figure \ref{fig:KSlogCP} that there always exist
threshold points when $P\neq0$. A more recent paper argued that this
divergent points of $C_{P}$ is associated to a second-order phase
transition\cite{Jing:2010cy}.

\begin{figure}[H]
\centering{}\includegraphics{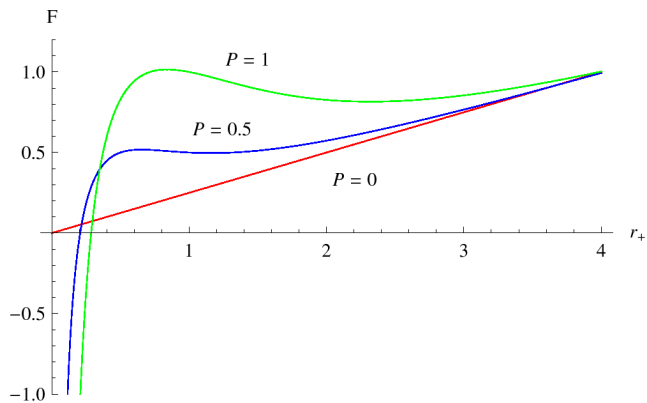}\caption{Free energy of KS black holes. \label{fig:KSlogF}}

\end{figure}

We also plot the free energy \begin{equation}
F=m-TS=\frac{4P^{4}+7P^{2}r_{+}^{2}+r_{+}^{4}+2\,\ln\left(r_{+}^{2}\right)\left(P^{4}-P^{2}r_{+}^{2}\right)}{4\left(2P^{2}r_{+}+r_{+}^{3}\right)}\quad,\end{equation}
 in Figure \ref{fig:KSlogF}. We see that the free energy for $P\neq0$
has a local maximum at $\xi_{1}=0$.

The above results can be interpreted as follow. The specific heat
with constant $P$ is positive for small radius KS black holes, while
negative for a larger one. This indicates that the small KS black
holes are thermodynamic stable, while the large KS black holes are
unstable. It is nothing out of ordinary for a asymptotically flat
black hole to have negative specific heat since the Schwarzschild
black hole in Einstein gravity is asymptotically flat spacetime, it
has negative specific heat and is unstable. However it is unusual
that the small KS black holes are stable. This conclusion is also
supported by Figure \ref{fig:KSlogF}, since the unstable large KS
black holes have positive free energy while the stable small KS black
holes can have negative free energy when the radius is small enough.
$P=\sqrt{\frac{\kappa^{2}}{32\mu^{2}}}\neq0$ is the crucial condition
for this interesting phase structure in KS black holes, which means
the high derivative term in the H-L gravity action comes into play,
and change the thermodynamic properties of asymptotically flat black
holes.

Other three $C$'s all can have both positive and negative values
in suitable parameter regions. The threshold points are at certain
value of $P$ and $r_{+}$ which satisfy the condition \begin{eqnarray*}
\xi_{2}=0 & \quad or\quad & \xi_{3}=0\quad.\end{eqnarray*}
The behaviors of $C_{\Phi}$,$\tilde{C_{T}}$, and $\tilde{C_{S}}$
will reveal thermodynamic stability in different ensembles, interesting
phase structures in the KS black holes can be found by examine this
quantities.

Thermodynamical metrics can be constructed as we did in Section \ref{sec:Charged-black-hole},
and it is straightforward to calculate Ricci scalars of thermodynamical
metrics. Considering the complexity of the expression of those Ricci
scalars, we only show the denominators. They are given by

\begin{eqnarray}
D(R^{\left(S\right)}) & = & \pi\xi_{2}^{2}\quad,\\
D(R^{\left(M\right)}) & = & (P-r_{+})^{2}(P+r_{+})^{2}\left(2P^{2}+r_{+}^{2}\right)\xi_{2}^{2}\quad,\\
D(R^{\left(F\right)}) & = & \xi_{1}^{2}\xi_{3}^{2}\quad.\end{eqnarray}
We see that all possible phase transitions correspond to curvature
singularities of certain thermodynamical metrics, which are consistent
to the framework proposed by \cite{liu2010thermodynamical}.

\section{Conclusion\label{sec:Conclusion}}

Black hole phase transitions have been extensively studied since the
discovery of Hawking-Page phase transition and Witten's interpretation
of the transition in the frame of AdS/CFT correspondence. However,
the lack of exact knowledge about the microscopic statistical framework
underlying black hole thermodynamic renders the issue of phase transitions
in black holes far from being completely settled. In spite of these
difficulties, some local and global stability properties were examined
in the constructed canonical ensemble or microcanonical ensemble.
Local stability in the canonical ensemble is equivalent to the positivity
of the heat capacity. However, the local stability is not sufficient
to ensure global stability, there do exist regions which are locally
stable but globally unstable, as we found in this paper.

In this paper, we have studied black hole phase transitions in (deformed)
Ho\v{r}ava-Lifshitz gravity though stability analyses, including
the uncharged/charged topological black hole and KS black hole. Some
interesting observations have been made. Compared to the Einstein
gravity, the phase structure of black holes in H-L gravity changed
dramatically, the stability of small radius black holes are essential
different. Pursuing it's deep reason of this will help us to gain
some information about how gravity works in small scale spacetime.

The framework proposed in \cite{liu2010thermodynamical} is also associated
with the local stability analyses to black hole phase transitions.
It is still unclear whether this framework can uncover some global
stability properties of black holes. We found a probable counter-example
to this framework. There is an infinite discontinuity at the specific
heat curve for charged black hole with hyperbolic event horizon in
H-L gravity. However, this discontinuity does not have a corresponding
curvature singularities of thermodynamical metrics. The violation
of the correspondence in H-L gravity may be due to the non-spherical
topology of their event horizons. We guess that the topology of the
event horizon and the dimension of spacetime may influence the validity
of the correspondence. One should also note that this is associated
with a local phase transition, not a global phase transition.

\begin{acknowledgments}
We thank Y.J.Du, J.L.Li, Q.Ma and Y.Q.Wang for useful discussions.
Chen would like to thank the organizer and the participants of the
advanced workshop, {}``Dark Energy and Fundamental Theory\textquotedblright{}
supported by the Special Fund for Theoretical Physics from the National
Natural Science Foundation of China with grant No. 10947203, for stimulating
discussions and comments. The research is supported by the NNSF of
China Grant No. 11075138, No. 10775116, 973 Program Grant No. 2005CB724508.
\end{acknowledgments}

\section*{Appendix }

For KS black hole, there is another idea in defining the mass and
entropy. One can treat $m$ as a mass parameter, and to use the one
quarter area entropy and ADM mass\cite{myung2010adm,wang912first},
In this appendix, we will adopt this idea to investigate the thermodynamic
properties of KS black holes. The expressions of entropy and ADM mass
are given by \begin{equation}
S=\frac{A}{4}=\pi r_{+}^{2}\quad,\end{equation}
\begin{equation}
M=\frac{r_{+}}{2}-\frac{3\,\tan^{-1}\left(\sqrt{\omega}r_{+}\right)}{4\sqrt{\omega}}+\frac{3\pi}{8\sqrt{\omega}}\quad.\end{equation}
The mass parameter $m$ is Eq.(\ref{eq:KSm}), and temperature formula
is the same as Eq.(\ref{eq:KSTem}), discussions about this can be
found in Section \ref{sec:KS-black-hole}.

The potential $V$ corresponding to the charge like parameter $P$
is computed as \begin{equation}
V=-\frac{3\left((1+\omega r_{+}^{2}\right)\tan^{-1}\left(\sqrt{\omega}r_{+}\right)-\sqrt{\omega}r_{+}}{2\sqrt{2}\left(1+\omega r_{+}^{2}\right)}+\frac{3\sqrt{2}\pi}{8}\quad,\end{equation}
which satisfy the Smarr law\begin{equation}
M=2TS+VP\quad,\end{equation}
and first law of thermodynamics\begin{equation}
dM=TdS+VdP\quad.\end{equation}

\begin{figure}[H]
\centering{}\includegraphics{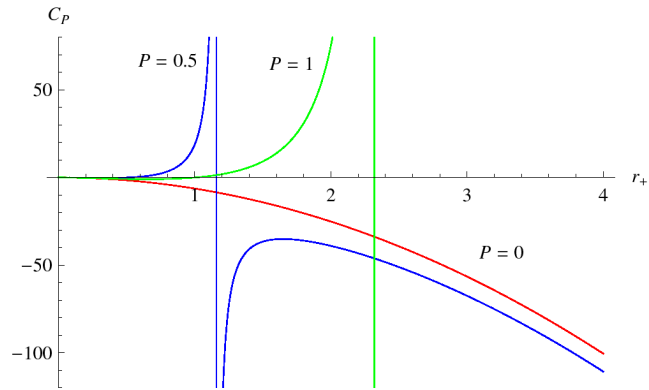}\caption{Heat capacity for KS black hole with area entropy.\label{fig:KSADMCP}}

\end{figure}

Furthermore, one can calculate the specific heats for constant $P$
or constant potential $V$, \begin{equation}
C_{P}=T\frac{\partial S}{\partial T}\biggl|_{P}=\frac{\partial M}{\partial T}\biggl|_{P}=\frac{2\pi r_{+}^{2}\left(-2P^{4}+P^{2}r_{+}^{2}+r_{+}^{4}\right)}{2P^{4}+5P^{2}r_{+}^{2}-r_{+}^{4}}\quad,\label{eq:cp for KS with area}\end{equation}
\begin{equation}
C_{V}=\frac{\partial M}{\partial T}\biggl|_{V}=\frac{\pi r_{+}\left(3\sqrt{2}\pi-6\sqrt{2}\,\tan^{-1}\left(\frac{\sqrt{\frac{1}{P^{2}}}r_{+}}{\sqrt{2}}\right)+4\sqrt{\frac{1}{P^{2}}}r_{+}\right)\left(2P^{2}+r_{+}^{2}\right)}{2\sqrt{\frac{1}{P^{2}}}\left(P^{2}-r_{+}^{2}\right)}\quad.\end{equation}
The capacitances at a fixed temperature or entropy are given respectively
by\begin{equation}
\tilde{C}_{T}=\frac{\partial P}{\partial V}\biggl|_{T}=\frac{\sqrt{\frac{1}{P^{2}}}P\left(2P^{2}+r_{+}^{2}\right)\left(2P^{4}+5P^{2}r_{+}^{2}-r_{+}^{4}\right)}{3r_{+}^{3}\left(P^{2}-r_{+}^{2}\right)}\quad,\end{equation}
\begin{equation}
\tilde{C}_{S}=\frac{\partial P}{\partial V}\biggl|_{S}=\frac{\sqrt{\frac{1}{P^{2}}}P\left(2P^{2}+r_{+}^{2}\right)^{2}}{3r_{+}^{3}}\quad.\end{equation}
The feature of the heat capacity can be seen from Figure \ref{fig:KSADMCP}.
$C_{P}$ also has a pole when $\xi_{1}=0$, which agrees with what
we get from Eq.(\ref{eq:cp for KS with log}). Also note that the
divergence of $C_{P}$ is associated with the vanishing of $\tilde{C}_{T}$.
But the other three $C$'s are very different from what we discussed
in Section \ref{sec:KS-black-hole}. $C_{V}$ and $\tilde{C_{T}}$
both have singularity at $P^{2}=r_{+}^{2}$, or $P^{2}=m^{2}$ where
the black hole is extremal, which may be the critical point for a
phase transition from extremal to non-extremal black hole. $\tilde{C}_{S}$
is not divergent in the physical region of parameters, and is not
relevant to KS black hole phase transition.

\begin{figure}[H]
\begin{centering}
\includegraphics{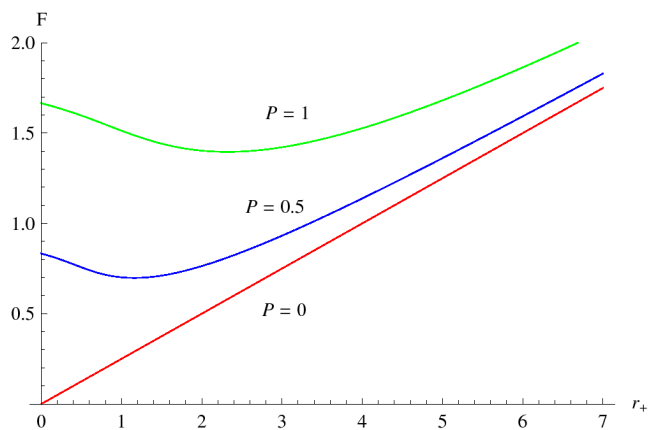}
\par\end{centering}

\caption{Free energy for KS black hole with are entropy.\label{fig:KSADMF}}

\end{figure}

The free energy\begin{eqnarray}
F & = & M-TS\nonumber \\
 & = & \frac{1}{8}\left(\frac{3\sqrt{2}\pi-6\sqrt{2}\,\tan^{-1}\left(\frac{\sqrt{\frac{1}{P^{2}}}r_{+}}{\sqrt{2}}\right)+4\sqrt{\frac{1}{P^{2}}}r_{+}}{\sqrt{\frac{1}{P^{2}}}}+\frac{2r_{+}\left(P^{2}-r_{+}^{2}\right)}{2P^{2}+r_{+}^{2}}\right)\nonumber \\
 & = & \frac{1}{8}\left(3\sqrt{2}\left|P\right|\left(\pi-2\,\tan^{-1}\left(\frac{r_{+}}{\sqrt{2}\left|P\right|}\right)\right)+4r_{+}+\frac{2r_{+}\left(P^{2}-r_{+}^{2}\right)}{2P^{2}+r_{+}^{2}}\right)\end{eqnarray}
is plotted in Figure \ref{fig:KSADMF}. %
The minimum of free energy corresponds to the divergent point of $C_{P}$,
and the maximum point of temperature.

The Ricci scalars of the the Ruppeiner and Weinhold metrics, and the
thermodynamical metric derived from the free energy are given by\begin{equation}
R^{(S)}=0\quad,\end{equation}
\begin{equation}
R^{(M)}=-\frac{2\left(2P^{4}+r_{+}^{4}\right)}{r_{+}\left(P^{2}-r_{+}^{2}\right)^{2}}\quad,\end{equation}
\begin{equation}
R^{(F)}=-\frac{4\left(2P^{2}+r_{+}^{2}\right)^{2}\left(2P^{4}+r_{+}^{4}\right)}{r_{+}\left(-2P^{4}-5P^{2}r_{+}^{2}+r_{+}^{4}\right)^{2}}\quad.\end{equation}
It is not surprise that the Ruppeiner curvature $R^{\left(S\right)}$
vanishes, since we have treating the KS black hole similar with the
RN black hole, while the Ruppeiner curvature for the latter is zero.

We see that the divergent point of $R^{\left(M\right)}$ corresponds
to the divergent point of $C_{V}$ and $\tilde{C}_{T}$, and that
the divergent point of $R^{\left(F\right)}$ correspond to the divergent
point of $C_{P}$. The results also agree with the analyzes in \cite{liu2010thermodynamical}.

\bibliographystyle{hunsrt}
\bibliography{bhphasetransitions}

\begin{thebibliography}{10}

\bibitem{Bekenstein1973}
Jacob~D. Bekenstein.
\newblock {Black holes and entropy}.
\newblock {\em Phys. Rev.}, D7:2333--2346, 1973.

\bibitem{Hawking1975}
S.~W. Hawking.
\newblock {Particle Creation by Black Holes}.
\newblock {\em Commun. Math. Phys.}, 43:199--220, 1975.

\bibitem{bardeen1973four}
JM~Bardeen, B.~Carter, and SW~Hawking.
\newblock {The four laws of black hole mechanics}.
\newblock {\em Communications in Mathematical Physics}, 31(2):161--170, 1973.

\bibitem{hawking1983thermodynamics}
SW~Hawking and D.N. Page.
\newblock {Thermodynamics of black holes in anti-de Sitter space}.
\newblock {\em Communications in Mathematical Physics}, 87(4):577--588, 1983.

\bibitem{witten1998anti}
E.~Witten.
\newblock {Anti-de Sitter space, thermal phase transition, and confinement in
  gauge theories}.
\newblock {\em Adv. Theor. Math. Phys.}, 2(3):505--532, 1998.

\bibitem{davies1989thermodynamic}
PCW Davies.
\newblock {Thermodynamic phase transitions of Kerr-Newman black holes in de
  Sitter space}.
\newblock {\em Classical and Quantum Gravity}, 6:1909, 1989.

\bibitem{chamblin1999charged}
A.~Chamblin, R.~Emparan, C.V. Johnson, and R.C. Myers.
\newblock {Charged AdS black holes and catastrophic holography}.
\newblock {\em Physical Review D}, 60(6):64018, 1999.

\bibitem{chamblin1999holography}
A.~Chamblin, R.~Emparan, C.V. Johnson, and R.C. Myers.
\newblock {Holography, thermodynamics, and fluctuations of charged AdS black
  holes}.
\newblock {\em Physical Review D}, 60(10):104026, 1999.

\bibitem{caldarelli2000thermodynamics}
M.M. Caldarelli, G.~Cognola, and D.~Klemm.
\newblock {Thermodynamics of Kerr-Newman-AdS black holes and conformal field
  theories}.
\newblock {\em Classical and Quantum Gravity}, 17:399, 2000.

\bibitem{carter2005thermodynamics}
B.M.N. Carter and I.P. Neupane.
\newblock {Thermodynamics and stability of higher dimensional rotating (Kerr-)
  AdS black holes}.
\newblock {\em Physical Review D}, 72(4):43534, 2005.

\bibitem{cai2002gauss}
R.G. Cai.
\newblock {Gauss-Bonnet black holes in AdS spaces}.
\newblock {\em Physical Review D}, 65(8):84014, 2002.

\bibitem{cai2007ricci}
R.G. Cai, S.P. Kim, and B.~Wang.
\newblock {Ricci flat black holes and Hawking-Page phase transition in
  Gauss-Bonnet gravity and dilaton gravity}.
\newblock {\em Physical Review D}, 76(2):24011, 2007.

\bibitem{myung2008phase}
Y.S. Myung.
\newblock {Phase transition for black holes with scalar hair and topological
  black holes}.
\newblock {\em Physics Letters B}, 663(1-2):111--117, 2008.

\bibitem{dey2007phase}
T.K. Dey, S.~Mukherji, S.~Mukhopadhyay, and S.~Sarkar.
\newblock {Phase transitions in higher derivative gravity}.
\newblock {\em Journal of High Energy Physics}, 04:014, 2007.

\bibitem{sheykhi2009thermodynamical}
A.~Sheykhi.
\newblock {THERMODYNAMICAL PROPERTIES OF TOPOLOGICAL BORN- INFELD-DILATON BLACK
  HOLES}.
\newblock {\em International Journal of Modern Physics D}, 18(1):25--42, 2009.

\bibitem{fernando2006thermodynamics}
S.~Fernando.
\newblock {Thermodynamics of Born-Infeld--anti-de Sitter black holes in the
  grand canonical ensemble}.
\newblock {\em Physical Review D}, 74(10):104032, 2006.

\bibitem{myung2008thermodynamics}
Y.S. Myung, Y.W. Kim, and Y.J. Park.
\newblock {Thermodynamics and phase transitions in the Born-Infeld-anti-de
  Sitter black holes}.
\newblock {\em Physical Review D}, 78(8):84002, 2008.

\bibitem{anninos2009thermodynamics}
D.~Anninos and G.~Pastras.
\newblock {Thermodynamics of the Maxwell-Gauss-Bonnet anti-de Sitter black hole
  with higher derivative gauge corrections}.
\newblock {\em Journal of High Energy Physics}, 07:030, 2009.

\bibitem{dehghani2009thermodynamic}
MH~Dehghani and R.~Pourhasan.
\newblock {Thermodynamic instability of black holes of third order Lovelock
  gravity}.
\newblock {\em Physical Review D}, 79(6):64015, 2009.

\bibitem{davies1977thermodynamic}
PCW Davies.
\newblock {The thermodynamic theory of black holes}.
\newblock {\em Proceedings of the Royal Society of London. Series A,
  Mathematical and Physical Sciences}, 353(1675):499--521, 1977.

\bibitem{davies1978thermodynamics}
PCW Davies.
\newblock {Thermodynamics of black holes}.
\newblock {\em Reports on Progress in Physics}, 41:1313, 1978.

\bibitem{birmingham1999topological}
D.~Birmingham.
\newblock {Topological black holes in anti-de Sitter space}.
\newblock {\em Classical and Quantum Gravity}, 16:1197, 1999.

\bibitem{koutsoumbas2008phase}
G.~Koutsoumbas, E.~Papantonopoulos, and G.~Siopsis.
\newblock {Phase transitions in charged topological-AdS black holes}.
\newblock {\em Journal of High Energy Physics}, 05:107, 2008.

\bibitem{weinhold1975}
Weinhold.F.
\newblock {\em J.Chem.Phys}, 63:2479, 1975.

\bibitem{ruppeiner1979thermodynamics}
G.~Ruppeiner.
\newblock {Thermodynamics: A Riemannian geometric model}.
\newblock {\em Physical Review A}, 20(4):1608--1613, 1979.

\bibitem{ruppeiner1995riemannian}
G.~Ruppeiner.
\newblock {Riemannian geometry in thermodynamic fluctuation theory}.
\newblock {\em Reviews of Modern Physics}, 67(3):605--659, 1995.

\bibitem{aman2003geometry}
J.E. {\AA}man, I.~Bengtsson, and N.~Pidokrajt.
\newblock {Geometry of black hole thermodynamics}.
\newblock {\em General Relativity and Gravitation}, 35(10):1733--1743, 2003.

\bibitem{shen2007thermodynamic}
J.~Shen, R.G. Cai, B.~Wang, and R.K. Su.
\newblock {Thermodynamic geometry and critical behavior of black holes}.
\newblock {\em International Journal of Modern Physics A}, 22(1):11--28, 2007.

\bibitem{ruppeiner2008thermodynamic}
G.~Ruppeiner.
\newblock {Thermodynamic curvature and phase transitions in Kerr-Newman black
  holes}.
\newblock {\em Physical Review D}, 78(2):24016, 2008.

\bibitem{sahay2010thermodynamic}
A.~Sahay, T.~Sarkar, and G.~Sengupta.
\newblock {Thermodynamic geometry and phase transitions in Kerr-Newman-AdS
  black holes}.
\newblock {\em Journal of High Energy Physics}, 04(4):118, 2010.

\bibitem{liu2010thermodynamical}
Haishan Liu, H.~Lu, Mingxing Luo, and Kai-Nan Shao.
\newblock {Thermodynamical Metrics and Black Hole Phase Transitions}.
\newblock {\em JHEP}, 12:054, 2010, 1008.4482.

\bibitem{Horava2009}
Petr Horava.
\newblock {Quantum Gravity at a Lifshitz Point}.
\newblock {\em Phys. Rev.}, D79:084008, 2009, 0901.3775.

\bibitem{Horava2009a}
Petr Horava.
\newblock {Membranes at Quantum Criticality}.
\newblock {\em JHEP}, 03:020, 2009, 0812.4287.

\bibitem{Horava2009b}
Petr Horava.
\newblock {Spectral Dimension of the Universe in Quantum Gravity at a Lifshitz
  Point}.
\newblock {\em Phys. Rev. Lett.}, 102:161301, 2009, 0902.3657.

\bibitem{Lu2009a}
H.~Lu, Jianwei Mei, and C.~N. Pope.
\newblock {Solutions to Horava Gravity}.
\newblock {\em Phys. Rev. Lett.}, 103:091301, 2009, 0904.1595.

\bibitem{cai2009topological}
R.G. Cai, L.M. Cao, and N.~Ohta.
\newblock {Topological black holes in Ho{\v{r}}ava-Lifshitz gravity}.
\newblock {\em Physical Review D}, 80(2):24003, 2009.

\bibitem{kehagias2009black}
A.~Kehagias and K.~Sfetsos.
\newblock {The black hole and FRW geometries of non-relativistic gravity}.
\newblock {\em Physics Letters B}, 678(1):123--126, 2009.

\bibitem{Ghodsi:2009zi}
Ahmad Ghodsi and Ehsan Hatefi.
\newblock {Extremal rotating solutions in Horava Gravity}.
\newblock {\em Phys.Rev.}, D81:044016, 2010, 0906.1237.

\bibitem{cai2009thermodynamics}
R.G. Cai, L.M. Cao, and N.~Ohta.
\newblock {Thermodynamics of black holes in Horava-Lifshitz gravity}.
\newblock {\em Physics Letters B}, 679(5):504--509, 2009.

\bibitem{biswas2010geometry}
R.~Biswas and S.~Chakraborty.
\newblock {Geometry of the thermodynamics of the black holes in Horava-Lifshitz
  gravity}.
\newblock {\em General Relativity and Gravitation}, 43:41, 2011.

\bibitem{wei2010thermodynamic}
S.W. Wei, Y.X. Liu, Y.Q. Wang, and H.~Guo.
\newblock {Thermodynamic Geometry of black hole in the deformed Horava-Lifshitz
  gravity}.
\newblock {\em Arxiv preprint arXiv:1002.1550}, 2010.

\bibitem{Carlip:2003ne}
Steven Carlip and S.~Vaidya.
\newblock {Phase transitions and critical behavior for charged black holes}.
\newblock {\em Class.Quant.Grav.}, 20:3827--3838, 2003, gr-qc/0306054.

\bibitem{Banerjee:2010da}
Rabin Banerjee, Sumit Ghosh, and Dibakar Roychowdhury.
\newblock {New type of phase transition in Reissner Nordstr\'om - AdS black
  hole and its thermodynamic geometry}.
\newblock {\em Phy. Lett. B}, 696:156, 2011.

\bibitem{myung2009thermodynamics}
Y.S. Myung.
\newblock {Thermodynamics of black holes in the deformed Horava-Lifshitz
  gravity}.
\newblock {\em Physics Letters B}, 678(1):127--130, 2009.

\bibitem{myung2010entropy}
Y.S. Myung.
\newblock {Entropy of black holes in the deformed Horava-Lifshitz gravity}.
\newblock {\em Physics Letters B}, 684(2-3):158--161, 2010.

\bibitem{castillo906entropy}
A.~Castillo and A.~Larranaga.
\newblock {Entropy for black holes in the deformed Horava-Lifshitz gravity}.
\newblock {\em arXiv}, 0906.4380.

\bibitem{myung2010adm}
Y.S. Myung.
\newblock {ADM mass and quasilocal energy of black hole in the deformed
  Horava-Lifshitz gravity}.
\newblock {\em Physics Letters B}, 685(4-5):318--324, 2010.

\bibitem{wang912first}
M.~Wang, J.~Jing, C.~Ding, and S.~Chen.
\newblock {First laws of thermodynamics in IR Modified Horava-Lifshitz
  gravity}.
\newblock {\em Phys. Lett. B}, 695:401, 2011.

\bibitem{Jing:2010cy}
Mengjie Wang Songbai Chen~Jiliang Jing.
\newblock {Second-order phase transition of Kehagias-Sfetsos black hole in
  deformed H\v{o}rava-Lifshitz gravity}.
\newblock {\em Phys. Lett.}, B695:401, 2010, 1012.0645.

\end{thebibliography}

\end{document}